\documentclass[prd,twocolumn,showpacs,preprintnumbers,amsmath,amssymb]{revtex4}
\usepackage{graphicx}
\usepackage{epsfig}
\usepackage{dcolumn}
\usepackage{bm}

\newcommand{\BR}{{\cal B}}
\newcommand{\eff}{\varepsilon}
\newcommand{\psip}{\psi(2S)}

\newcommand{\chicJ}{\chi_{cJ}}
\newcommand{\jpsi}{J/\psi}
\newcommand{\etac}{\eta_{c}}
\newcommand{\EE}{e^+e^-}
\newcommand{\MM}{\mu^+\mu^-}

\newcommand{\pip}{\pi^+}
\newcommand{\pim}{\pi^-}
\newcommand{\piz}{\pi^0}
\newcommand{\pp}{\pi^+\pi^-}
\newcommand{\kap}{K^+}
\newcommand{\kam}{K^-}
\newcommand{\ks}{K^0_S}
\newcommand{\kstar}{K^{\ast}}
\newcommand{\kstarz}{K^{\ast0}}
\newcommand{\kstarzb}{\overline{K}^{\ast0}}

\newcommand{\ppb}{p\overline{p}}

\newcommand{\ppjpsi}{\pi^+\pi^- J/\psi}
\newcommand{\ra}{\rightarrow}
\newcommand{\jpsito}{J/\psi \rightarrow }

\newcommand{\psipto}{\psi(2S) \rightarrow }

\newcommand{\rhopi}{\rho\pi}

\newcommand{\beq}{\begin{equation}}
\newcommand{\eeq}{\end{equation}}
\newcommand{\bfg}{\begin{figure}}
\newcommand{\efg}{\end{figure}}
\newcommand{\bitm}{\begin{itemize}}
\newcommand{\eitm}{\end{itemize}}
\newcommand{\bnum}{\begin{enumerate}}
\newcommand{\enum}{\end{enumerate}}
\newcommand{\btbl}{\begin{table}}
\newcommand{\etbl}{\end{table}}
\newcommand{\btbu}{\begin{tabular}}
\newcommand{\etbu}{\end{tabular}}

\newcommand{\mev}{~\mbox{MeV}}
\newcommand{\gev}{~\mbox{GeV}}
\newcommand{\kskp}{K^0_S K^+ \pi^- + c.c.}
\def\etal{{\it et al.}}
\begin{document}

\title{\boldmath $\psip$ Radiative Decay Measurements }

\author{
M.~Ablikim$^{1}$,              J.~Z.~Bai$^{1}$,   Y.~Bai$^{1}$,
Y.~Ban$^{12}$, X.~Cai$^{1}$,                  H.~F.~Chen$^{17}$,
H.~S.~Chen$^{1}$,              H.~X.~Chen$^{1}$, J.~C.~Chen$^{1}$,
Jin~Chen$^{1}$
% ...
\\
\vspace{0.2cm}
(BES Collaboration)\\
\vspace{0.2cm} {\it
$^{1}$ Institute of High Energy Physics, Beijing 100049, People's Republic of China\\
% ...
$^{12}$ Peking University, Beijing 100871, People's Republic of China\\
% ...
$^{17}$ University of Science and Technology of China, Hefei 230026,
People's Republic of China\\
% ...
} }

\date{\today}

\begin{abstract}
  Using a sample of 14 million $\psip$ events collected with the BESII
  detector, branching fractions or upper limits on the branching
  fractions of $\psip$ decays into $\gamma\ppb$,
  $\gamma 2(\pip\pim)$, $\gamma \kskp$, $\gamma \kap\kam\pip\pim$,
  $\gamma\kstarz\kam\pip+c.c.$, $\gamma \kstarz \kstarzb$,
  $\gamma\pip\pim\ppb$, $\gamma2(\kap\kam)$, $\gamma3(\pip\pim)$ and
  $\gamma2(\pip\pim)\kap\kam$
  with hadron invariant mass less than 2.9$\gev/c^2$ are reported. We
  also report branching fractions of $\psip$ decays into $\gamma\ppb\piz$,
  $\ppb\piz\piz$, $2(\pip\pim)\piz$, $\omega\pp$, $\omega f_2(1270)$,
  $b_1^\pm\pi^\mp$ , $\piz\kskp$, $K^\pm\rho^\mp\ks$,
  $\piz2(\pip\pim)\kap\kam$ and $\gamma\piz2(\pip\pim)\kap\kam$.
\end{abstract}

\maketitle

\section{Introduction}
Besides the conventional meson and baryon states, QCD (Quantum
Chromodynamics) also predicts a rich spectrum of so-called QCD
exotics, among them glueballs ($gg$), hybrids ($q\bar{q}g$), and
four-quark states ($qq\bar{q}\bar{q}$) in the 1.0 to 2.5 $\gev/c^2$
mass region.  Therefore, the search for evidence of these exotic
states plays an important role in testing QCD.  Radiative decays of
quarkonium are expected to be a good place to look for glueballs,
and there have been many studies performed in $\jpsi$ radiative
decays~\cite{Jdecay, QWG}, while such studies have been limited in
$\psip$ radiative decays due to low statistics in previous
experiments~\cite{PDG, QWG}. Radiative decays of $\psip$ to light
hadrons are expected at about 1\% of its total decay
width~\cite{PRD-wangp}. However, the previously measured channels
only sum up to 0.05\%~\cite{PDG}.

In this paper we present measurements of $\psip$ decays into
  $\gamma\ppb$,  $\gamma 2(\pip\pim)$, $\gamma
  \kskp$, $\gamma \kap\kam\pip\pim$, $\gamma\kstarz\kam\pip+c.c.$,
  $\gamma \kstarz \kstarzb$, $\gamma\pip\pim\ppb$,
  $\gamma2(\kap\kam)$, $\gamma3(\pip\pim)$ and
  $\gamma2(\pip\pim)\kap\kam$
  with the invariant mass of the hadrons ($m_{hs}$) in each final
  state less than 2.9$\gev/c^2$. Measurements of $\psip$ decays into
  $\gamma\ppb\piz$,
  $\ppb\piz\piz$, $2(\pip\pim)\piz$, $\omega\pp$, $\omega f_2(1270)$,
  $b_1^\pm\pi^\mp$ , $\piz\kskp$, $K^\pm\rho^\mp\ks$,
  $\piz2(\pip\pim)\kap\kam$  and $\gamma\piz2(\pip\pim)\kap\kam$ are also presented and are used for the
  background analysis. Many of the above measurements have been
  published previously~\cite{prlrad}; this paper provides more
  detailed information.

\section{BES detector}
BES is a conventional solenoidal magnetic detector that is described
in detail in Refs.~\cite{bes,bes2}. A 12-layer vertex chamber (VTC)
surrounding the beam pipe provides trigger and track information. A
forty-layer main drift chamber (MDC), located radially outside the
VTC, provides trajectory and energy loss ($dE/dx$) information for
charged tracks over $85\%$ of the total solid angle.  The momentum
resolution is $\sigma _p/p = 1.78\% \sqrt{1+p^2}$ ($p$ in $\gev/c$),
and the $dE/dx$ resolution for hadron tracks is $\sim 8\%$. An array
of 48 scintillation counters surrounding the MDC measures the
time-of-flight (TOF) of charged tracks with a resolution of $\sim
200$ ps for hadrons.  Radially outside the TOF system is a 12
radiation length, lead-gas barrel shower counter (BSC).  This
measures the energies of electrons and photons over $\sim 80\%$ of
the total solid angle with an energy resolution of
$\sigma_E/E=21\%/\sqrt{E}$ ($E$ in GeV). Outside of the solenoidal
coil, which provides a 0.4~Tesla magnetic field over the tracking
volume, is an iron flux return that is instrumented with three
double layers of counters that identify muons of momentum greater
than 0.5$\gev/c$.

The data sample used in this analysis was taken with the BESII
detector at the BEPC storage ring at an energy of $\sqrt{s}=
3.686\gev$.  The number of $\psip$ events is ($14.0\pm 0.6)\times
10^6$, determined from inclusive hadronic events~\cite{pspscan},
and corresponds to a luminosity of $\mathcal{L}_{3.686} =
(19.72\pm0.86)~\hbox{pb}^{-1}$~\cite{lum}, measured with large angle
Bhabha events.  Continuum data, used for background studies, were
taken at $\sqrt{s} = 3.650\gev$ with a luminosity of
$\mathcal{L}_{3.650}=(6.42\pm0.24)~\hbox{pb}^{-1}$~\cite{lum}.  The
ratio of the two luminosities is
$\mathcal{L}_{3.686}/\mathcal{L}_{3.650} = 3.07\pm0.09$.

Monte Carlo (MC) simulations are used for the determination of mass
resolutions and detection efficiencies, as well as background
studies. The simulation of the BESII detector is GEANT3 based, where
the interactions of particles with the detector material are
simulated. Reasonable agreement between data and Monte Carlo
simulation is observed~\cite{simbes} in various channels such as $\EE
\ra (\gamma)\EE$, $\EE\ra (\gamma)\MM$, $\jpsito \ppb$ and $\psipto
\ppjpsi$, $\jpsito \ell^+\ell^-$ $(\ell=e,\mu)$. An inclusive $\psip$
decay MC sample of the same size as the $\psip$ sample is generated by
LUNDCHARM~\cite{chenjc} and used to estimate backgrounds,

\section{Event selection \label{sel}}

A neutral cluster is taken as a photon candidate when the following
conditions are satisfied: the energy deposited in the BSC is greater
than 50$\mev$; the first hit is in the beginning six radiation lengths;
the angle between the cluster and the nearest charged track is greater
than $15^{\circ}$; and the difference between the angle of the cluster
development direction in the BSC and the photon emission direction is
less than $37^{\circ}$.

Each charged track is required to be well fitted to a three-dimensional
helix, be in the polar angle region $|\cos\theta|<0.8$ in the MDC, and
have a transverse momentum greater than 70$\mev/c$.  The particle
identification chi-squared,
$\chi^2_{PID}(i)$, is calculated based on the $dE/dx$ and TOF
measurements with the following definition
$$\chi^2_{PID}(i) = \chi^2_{dE/dx}(i) + \chi^2_{TOF}(i).$$ For all
analyzed decay channels, the final states of the candidate events must
have the correct number of charged tracks with net charge zero.

If there is more than one photon candidate in an event, the
candidate with the largest energy deposit in the BSC is taken as the
radiative photon in the event, and a four-constraint kinematic fit
(4C-fit) is performed.  The combined confidence level,
$prob(\chi^2_{comb},
ndf)=\frac{1}{\sqrt{2^{ndf}}\Gamma({ndf}/2)}\int_{\chi^2_{comb}}^{\infty}e^{-\frac{t}{2}}t^{\frac{{ndf}}{2}-1}dt$,
is required to be greater than 1\%, where $ndf$ is the number of
degrees of freedom and $\chi^2_{comb}$ is defined as the sum of the
$\chi^2$ of the kinematic fit ($\chi^2_{4C}$) and $\chi_{PID}^2$:
$\chi^2_{comb}=\chi_{4C}^2 + \sum\limits_{i} \chi_{PID}^2(i)$, where
$i$ runs over all charged tracks.

For each decay mode, $m_{hs}$ is required to be less than
2.9$\gev/c^2$ to exclude $\psip$ radiative transitions into other
charmonium states, such as $\chicJ$, $\jpsi$, and $\etac$.  To
remove background from charged particle misidentification, the value
of $\chi^2_{comb}$ for $\psip \to \gamma + hs$ is required to be
less than those for $\psip$ decays into background channels $\gamma
+ hs'$, where $hs'$ has the same number of charged tracks as $hs$,
but different particle types for the charged tracks.  If there is
potential background from $\psipto\ppjpsi$, it is largely suppressed
by applying $|m^{\pip\pim}_{recoil}-m_{J/\psi}|>0.05\gev/c^2$, where
$m^{\pip\pim}_{recoil}$ is the mass recoiling from each possible
$\pip\pim$ pair. Possible background from $\psipto\ks+X$ is removed
by requiring that the invariant mass of $\pip\pim$ is outside the
$\ks$ mass region ($|m_{\pi\pi}-m_{\ks}|>0.04\gev/c^2$).


\section{Backgrounds and fitting procedure \label{bkgs}}
In our analyses, the backgrounds for each $\psi(2S) \to \gamma + hs$
decay mode fall into three classes: (1) continuum background,
estimated using the continuum data; (2) multi-photon backgrounds,
e.g. $\psip\to \pi^0+hadrons$, $3\gamma+hadrons$, etc., where
$hadrons$ have the same charged tracks as the signal final state,
estimated with MC simulation and normalized according to their
branching fractions; and (3) other backgrounds, estimated using an
inclusive $\psip$ MC sample of 14 million events~\cite{chenjc}.
Multi-photon backgrounds are dominant; continuum background and
other backgrounds including contamination between the channels studied
are lower.  The observed $\chi^2_{4C}$ distributions include both
signal events and all of these backgrounds.

\begin{figure}[htbp] \centering
\includegraphics[width=0.45\textwidth]{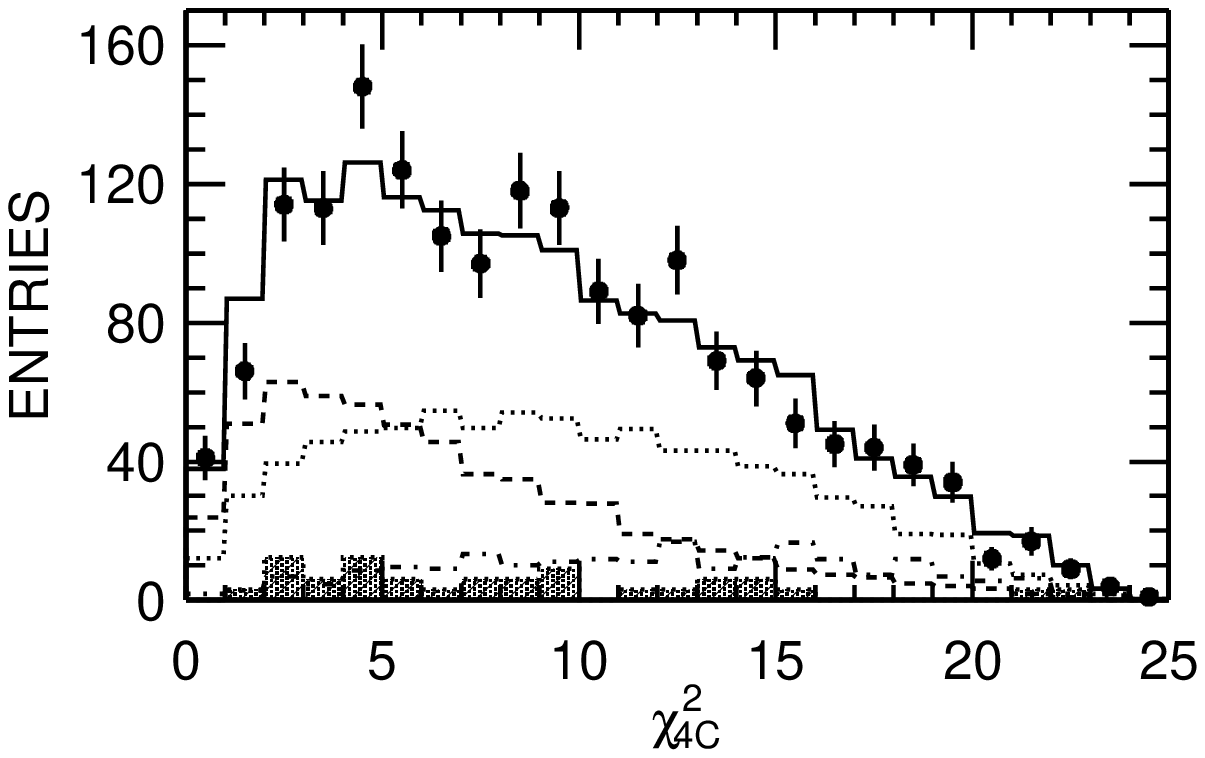}
\caption{ \label{chisqfit} The fitted $\chi^2_{4C}$ distribution for
$\psipto\gamma 2(\pip\pim)$ candidate events. The dots with error bars
are data. The solid line is the fitted result, which is the sum of the four
components: signal events (dashed line), MC simulated multi-photon
backgrounds (dotted line), continuum (hatched histogram), and
other backgrounds (dot-dashed line).}
\end{figure}

The number of signal events for most radiative decay channels is
extracted by fitting the observed $\chi^2_{4C}$ distributions with
those of the signal and background channels~\cite{fitchi2}, {\it i.e.}
$\chi^2_{obs}=w_s\chi^2_{sig}+\sum_{w_{b_i}}w_{b_i}\chi^2_{bg}$, where
$w_s$ and $w_{b_i}$ are the weights of the signal and the background
decays, respectively. As an example, Fig.~\ref{chisqfit} shows the
observed $\chi^2$ distribution for $\psipto\gamma 2(\pip\pim)$,
together with the $\chi^2$ distributions for the signal, multi-photon,
continuum, and other background channels, as well as the final fit. In
the fit, the weights of the multi-photon backgrounds and the continuum
backgrounds ($w_b$) are fixed to be the normalization factors, but the
weights of the signal ($w_s$) and the other backgrounds ($w_b$) are
free.  With this method, the number of signal events is extracted
for each radiative decay mode for $m_{hs}<2.9\gev/c^2$.

\section{Event analysis}

\subsection{\boldmath $\psipto\gamma\ppb$}
The major backgrounds to $\psipto\gamma\ppb$ come from
the channels $\psipto\piz\piz\ppb$, $\gamma\piz\ppb$, and $\piz\ppb$.
In order to estimate these backgrounds,
we first measure their branching fractions.

\subsubsection{Background estimation}
\begin{figure}
\begin{center}
\includegraphics[width=0.52\textwidth]{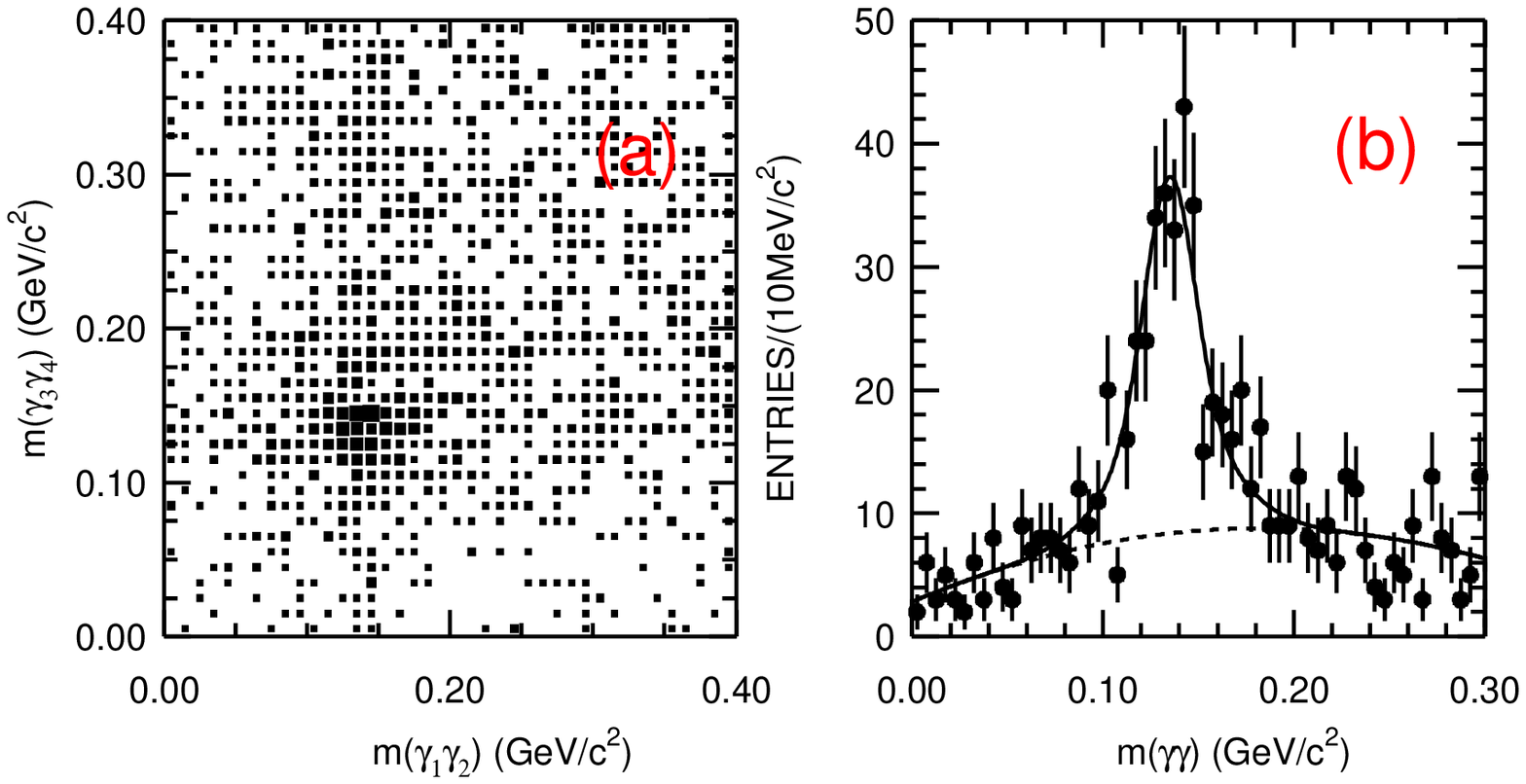}
\caption{\label{ppb2piz} (a) Scatter plot of $\gamma_1\gamma_2$ versus
  $\gamma_3\gamma_4$ invariant mass for $\psipto\piz\piz\ppb$
  candidate events. (b) Invariant mass distribution of
  $\gamma_1\gamma_2$ after requiring
  $|m_{\gamma_3\gamma_4}-m_{\pi^o}|<0.03\gev/c^2$.  The data are fitted
  with a double Gaussian function and a second order background polynomial. }
\end{center}
\end{figure}

For $\psipto\piz\piz\ppb$, candidate events must have two charged
tracks and four good photons. To reject background from
$\psipto\piz\piz\jpsi, \jpsito\ppb$ events, the $\ppb$ invariant mass
is required to satisfy
$|m_{\ppb}-m_{J/\psi}|>0.1\gev/c^2$. Figure~\ref{ppb2piz} (a) shows
the scatter plot of $\gamma_1\gamma_2$ versus $\gamma_3\gamma_4$
invariant mass for events after selection, where $\gamma_1\gamma_2$
and $\gamma_3\gamma_4$ are formed from all possible combinations of
the four photon candidates.  The cluster of events shows a clear
$\piz\piz$ pair signal.  Figure~\ref{ppb2piz} (b) shows the
${\gamma_1\gamma_2}$ invariant mass after requiring the other two
photons be consistent with being a $\pi^0$
($|m_{\gamma_3\gamma_4}-m_{\pi^o}|<0.03\gev/c^2$).  A fit to the peak
is performed with a double Gaussian function, where the parameters are
determined from MC simulation, plus a second order polynomial to
describe the smooth background. The number of events in the peak
determined from the fit is $254\pm24$, and the background
contamination to the peak is estimated to be $51\pm11$ from the
$m_{\gamma_3\gamma_4}$ sidebands: (0.0, 0.06) and (0.2, 0.26)
$\gev/c^2$.  No significant resonance is observed in the mass
distributions of any possible combination of particles in this final
state, indicating no significant intermediate processes in the
observed $\psipto\piz\piz\ppb$ candidate events. Therefore the
detection efficiency for $\psipto\piz\piz\ppb$ is determined to be
8.3\% using a phase space generator, and the branching fraction is
determined to be
\begin{equation}
\BR(\psipto\piz\piz\ppb)
=(1.75\pm0.21)\times10^{-4}, \nonumber
\end{equation}
where the error is statistical.

For $\psipto\gamma\piz\ppb$, candidate events must have two charged tracks
and three good photons. To reject background from
$\psipto\piz\piz\jpsi, \jpsito\ppb$ events, the $\ppb$ invariant mass
is required to satisfy $|m_{\ppb}-m_{J/\psi}|>0.1\gev/c^2$.

\begin{figure}
\begin{center}
\includegraphics[width=0.52\textwidth]{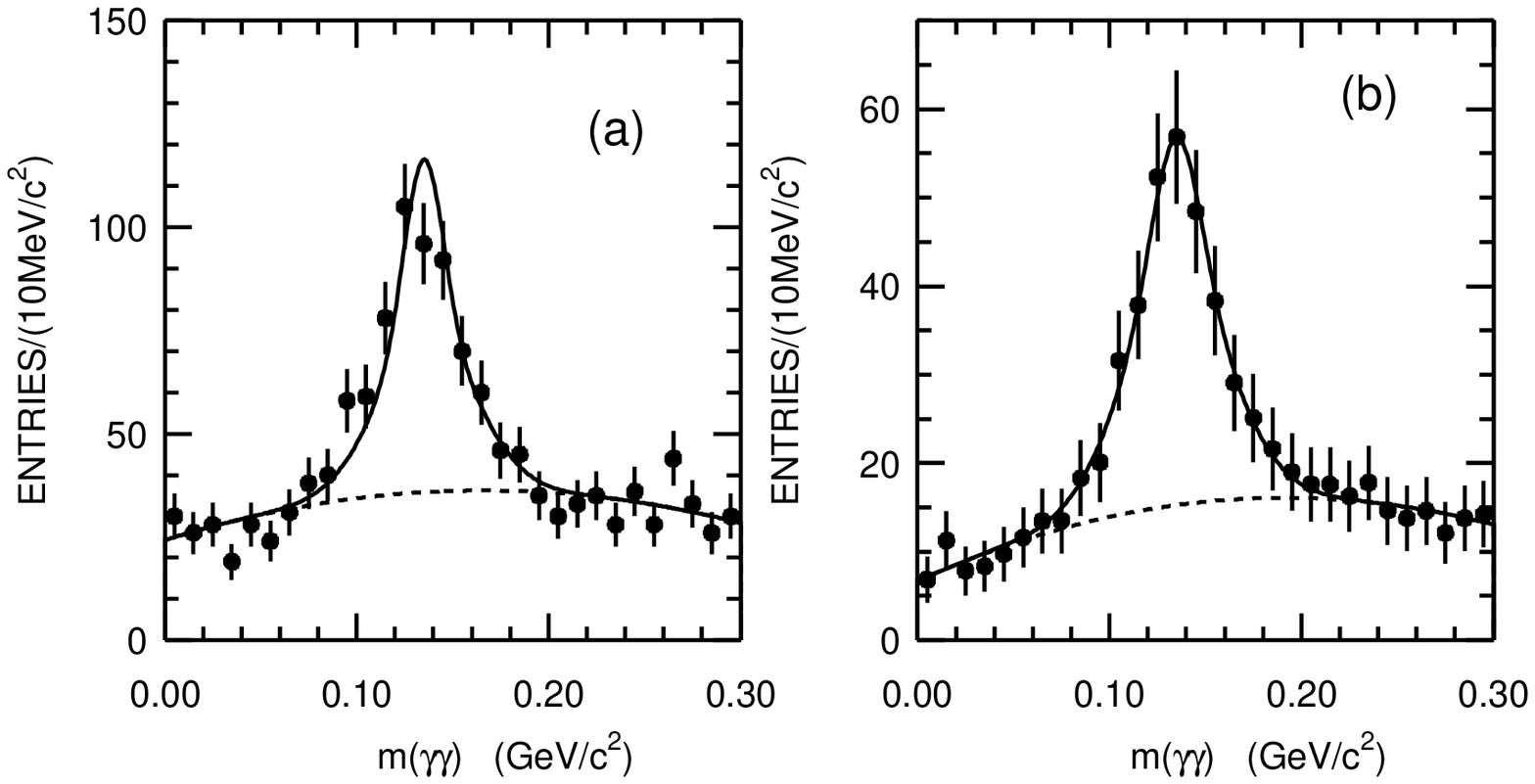}
\caption{\label{gpizppb} Distributions of $\gamma\gamma$ invariant
  mass for $\psipto\gamma\gamma\gamma\ppb$ candidate events fitted
  with a double Gaussian function and a second order background
  polynomial. (a) signal and (b) backgrounds including continuum,
  known simulated backgrounds ($\psipto\piz\piz\ppb$ and $\piz\ppb$),
  and other unknown backgrounds estimated from the $\psi(2S)$
  inclusive MC sample.}
\end{center}
\end{figure}

Figure~\ref{gpizppb} (a) shows the $\gamma\gamma$ invariant mass
distribution for $\psipto\gamma\gamma\gamma\ppb$ candidate events,
where $\gamma\gamma$ is any possible combination among the three
photon candidates. There is a clear $\piz$ signal.  The distribution
is fitted with a double Gaussian function with parameters determined
from MC simulation plus a second order polynomial for the background.
The number of events determined from the fit is $345\pm33$.

Background studies indicate that the main contamination to the $\piz$
signal comes from $\psipto\piz\piz\ppb$ and $\piz\ppb$; other
backgrounds only contribute a smooth background.  All possible
backgrounds, including continuum, known simulated backgrounds
($\psipto\piz\piz\ppb$ and $\piz\ppb$), and other unknown backgrounds
estimated from the $\psi(2S)$ inclusive MC sample, are combined in
Fig.~\ref{gpizppb} (b).  Fitting this distribution in the same way as
in Fig.~\ref{gpizppb} (a), the number of peaking background events is
estimated to be $219\pm18$.

Just as for $\psipto\piz\piz\ppb$, no
significant intermediate process is observed in
$\psipto\gamma\piz\ppb$. The efficiency is determined to be 8.94\%
with a phase space generator, and the branching fraction is calculated
to be
\begin{equation}
\BR(\psipto\gamma\piz\ppb) = (1.0\pm0.3)\times 10^{-4}, \nonumber
\end{equation}
where the  error is statistical.

For $\psipto\piz\ppb$, there is an earlier measurement from
BESII~\cite{ppbpi0-bes2}.  We reanalyze this channel in the same
way, and then extract the $m_{\ppb}$ distribution and estimate its
contamination to $\psipto\gamma\ppb$.

Candidate events are required to have two charged tracks
and two good photons. The probability of
the 4C-fit must be greater than 1\%, and the probability of the 4C-fit
for the $\psipto\gamma \gamma \ppb$ hypothesis must be greater than that for
$\psipto\gamma \gamma \kap\kam$. To reject background from
$\psipto\piz\piz\jpsi, \jpsito\ppb$ events, the invariant mass of $\ppb$
is required to be: $|m_{\ppb}-m_{J/\psi}|>0.02\gev/c^2$.

 A fit to the $m_{\gamma\gamma}$ distribution is performed with
 a double Gaussian function with parameters determined from MC
 simulation plus a second order polynomial for the background
 for $\psipto\gamma\gamma\ppb$ candidate events.
The number of events determined from the fit is $266\pm20$, and the
detection efficiency is 14.8\%. The branching fraction is determined
to be:
\begin{equation}
\BR(\psipto\piz\ppb) = (13.0\pm1.0)\times10^{-6}, \nonumber
\end{equation}
where the  error is statistical. This measurement agrees well with
the previous BESII result of
($13.2\pm1.0\pm1.5)\times10^{-6}$~\cite{ppbpi0-bes2}.

\subsubsection{Signal Analysis}
\begin{figure}[htbp] \centering
\includegraphics[width=0.45\textwidth]{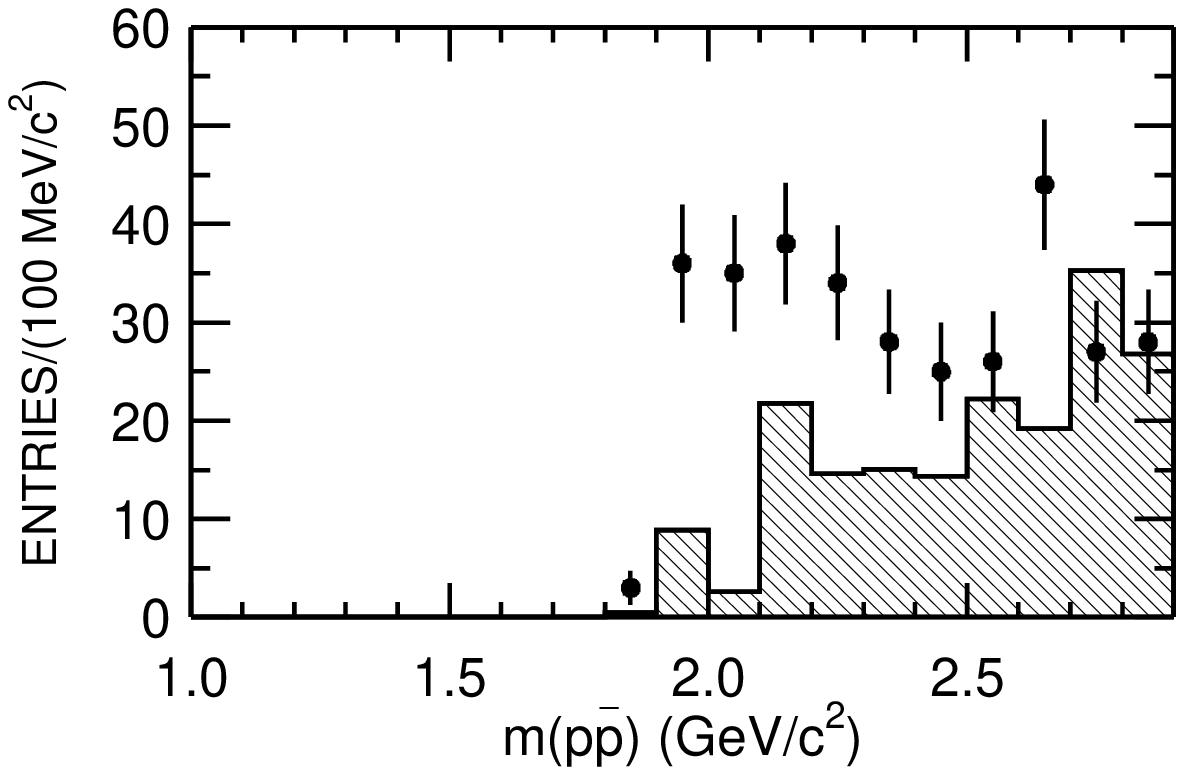}
\caption{\label{mppb0}The $\ppb$ invariant mass distribution for
$\psipto\gamma\ppb$ candidate events (dots with error bars). The
shaded histogram is the sum of all backgrounds, including continuum,
known simulated backgrounds ($\psi(2S) \to \pi^0 \pi^0 p \bar{p}$,
$\gamma \pi^0p\bar{p}$, and $\pi^0p\bar{p}$), and other unknown
backgrounds estimated from the $\psi(2S)$ inclusive MC sample.}
\end{figure}

For $\psipto\gamma\ppb$, 329 events are observed after the event
selection described in Section~\ref{sel}; the $\ppb$ invariant mass
distribution is shown in Fig.~\ref{mppb0}. After subtracting the
normalized major backgrounds, $\psi(2S) \to \pi^0 \pi^0 p \bar{p}$,
$\gamma \pi^0p\bar{p}$, and $\pi^0p\bar{p}$, the number of signal
events is $142\pm18$. The detection efficiency determined from MC
simulation is 35.3\%, and the branching fraction for this process is
determined to be:
\begin{equation}
\BR(\psipto\gamma \ppb) = (2.9\pm0.4)\times 10^{-5}, \nonumber
\end{equation}
where the error is statistical.

\begin{figure}[htbp] \centering
\includegraphics[width=0.47\textwidth]{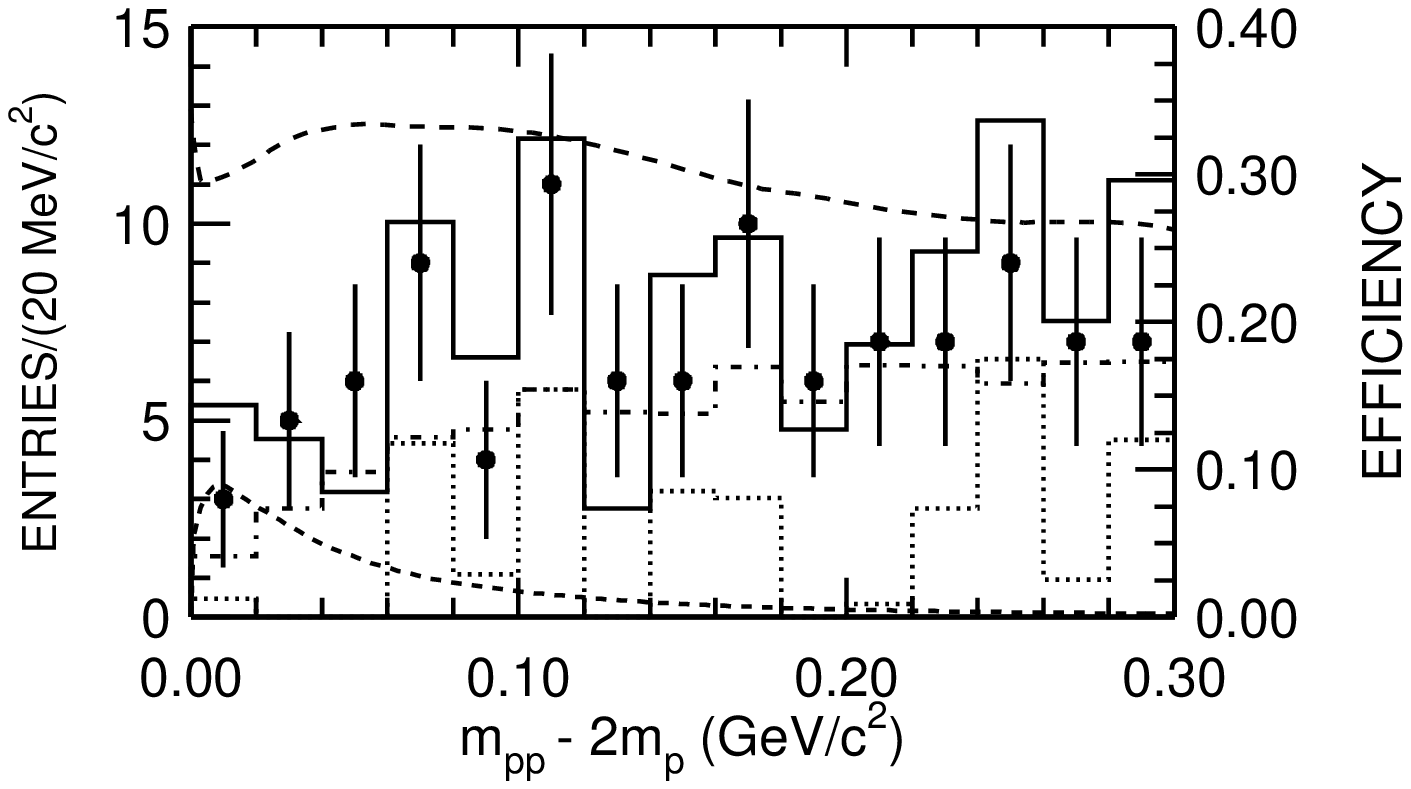}
\caption{\label{mppb} The fit to the $m_{\ppb}- 2m_p$ distribution of
$\psipto\gamma\ppb$ candidate events.  The solid histogram is the fit
result, the lower dashed line is the $X$ resonance shape, the
dash-dotted histogram is the shape for $\psi(2S) \to \gamma p\bar{p}$
phase-space, the dotted histogram is the measured background channels
 ($\psi(2S) \to \pi^0 \pi^0 p \bar{p}$,
$\gamma \pi^0p\bar{p}$, and $\pi^0p\bar{p}$), and the
top dashed line is the efficiency curve.}
\end{figure}

 There is an excess of events between $\ppb$
threshold and $2.5\gev/c^2$, but no significant narrow structure due
to the $X(1859)$, that was observed in
$\jpsito\gamma\ppb$~\cite{jpsi-gppb}. A fit to the mass spectrum (see
Fig.~\ref{mppb}) with an acceptance-weighted $S$-wave Breit-Wigner for
the $X$ resonance (with mass and width fixed to $1859\mev/c^2$ and
$30\mev/c^2$, respectively), together with the normalized MC
histograms for the above measured background channels ($\psi(2S) \to
\pi^0 \pi^0 p \bar{p}$, $\gamma \pi^0p\bar{p}$, and $\pi^0p\bar{p}$)
and the histogram from $\psi(2S) \to \gamma p\bar{p}$ phase
space~\cite{massres}, yields $11.7\pm 6.7$ events with a statistical
significance of 2.0 $\sigma$. The upper limit on the branching
fraction is determined to be
\begin{equation}
\BR(\psipto\gamma X(1859)\ra \gamma\ppb)<5.4\times 10^{-6} \nonumber
\end{equation}
at the 90\% C.L.

\subsection{\boldmath $\psipto\gamma2(\pip\pim)$}
For $\psipto\gamma2(\pip\pim)$, the main background comes from
$\psipto 2(\pip\pim)\piz$, so we first measure $\psipto2(\pip\pim)\piz$
in order to be able to estimate its contamination to
$\psipto\gamma2(\pip\pim)$.

\subsubsection{Background analysis}
For $\psipto\piz2(\pip\pim)$, candidate events are required to have
four charged tracks and two good photons.  The probability of the
4C-fit must be greater than 1\%, the $\chi^2_{comb}$ probability for
the $\psipto\piz2(\pip\pim)$ hypothesis must be greater than those of
$\psipto\piz\kap\kam\pip\pim$ and $\psipto\piz\pip\pim\ppb$, and the sum
of the momentum of any $\pip$ and $\pim$ pairs must be greater than
$550\mev/c$ to reject contamination from $\psipto\ppjpsi$ events.

After the above selection, a clear $\piz$ signal can be seen in the
$\gamma\gamma$ invariant mass distribution of $\psipto
2(\pip\pim)\gamma\gamma$ candidates. After subtracting backgrounds,
such as $\psi(2S) \to \gamma \chi_{c0}, \chi_{c0} \to
2(\pi^+\pi^-)$, $\psi(2S) \to \pi^0\pi^0 J/\psi, J/\psi \to
2(\pi^+\pi^-)$, etc., in the $\gamma\gamma$ invariant mass spectrum,
the distribution is fitted with a $\piz$ signal shape determined
with MC simulation plus a second order polynomial for the other
remaining backgrounds, and the number of $\piz$ signal events is
$2173\pm53$.
 The detection
efficiency is determined to be 6.32\% taking into consideration the
significant intermediate states such as $\omega\pip\pim$, $\omega
f_2(1270)$, and $b_1^\pm\pi^\mp$ described below.

\begin{figure}[htbp]\centering
\includegraphics[width=0.5\textwidth]{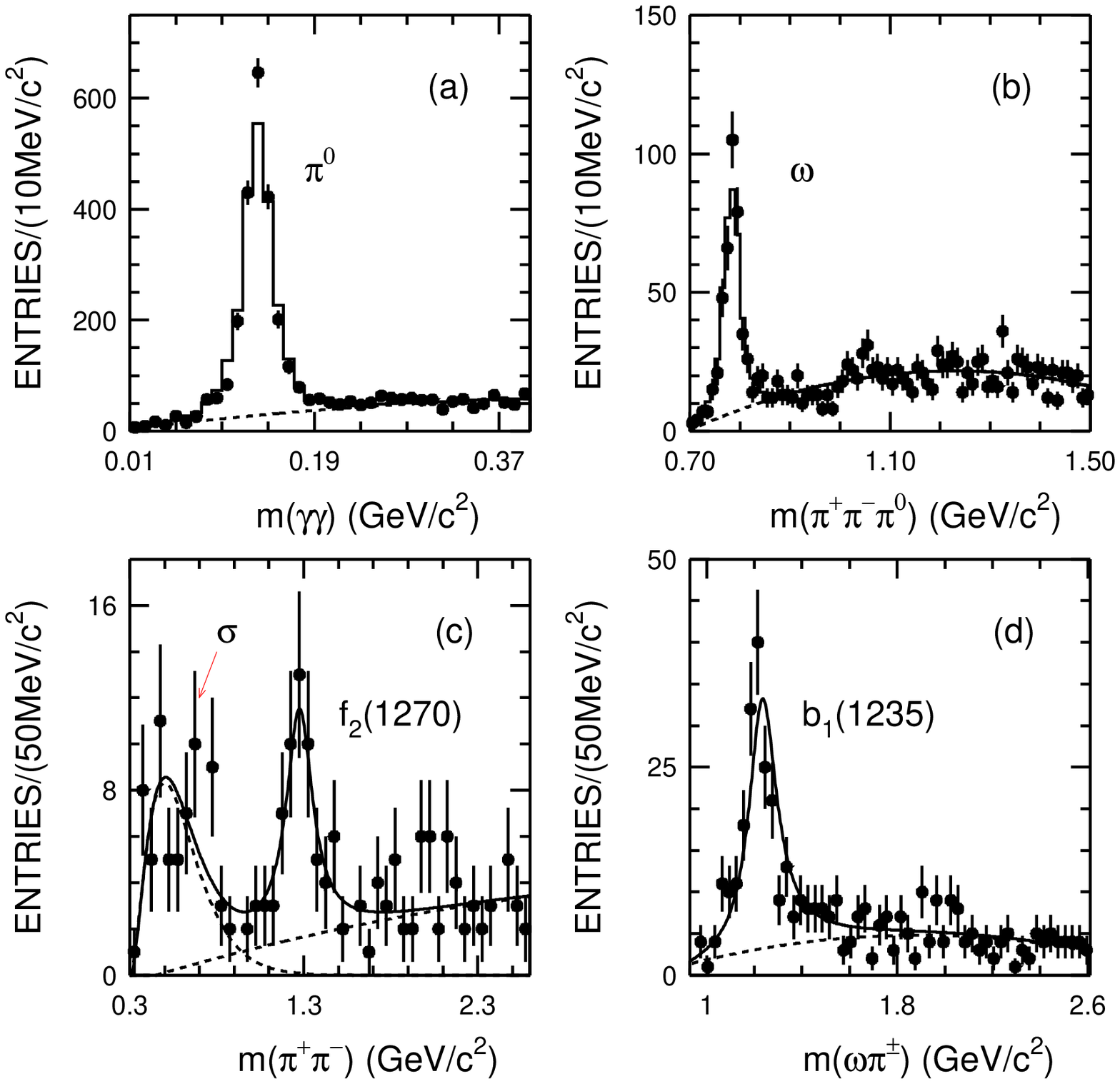}
\caption{\label{m5piall} Invariant mass
distributions with fits for $\psipto\piz2(\pip\pim)$ candidates, where
dots with error bars are data, and the solid histograms and curves
are the fit results. (a) $\gamma\gamma$; (b) $\pip\pim\piz$
with $|m_{\gamma\gamma}-m_{\pi^o}|<0.03\gev/c^2$; (c) $\pip\pim$ with
$|m_{\pip\pim\piz}-m_{\omega}|<0.05\gev/c^2$ and $b_1^\pm\pi^\mp$ events rejected;
and (d) $\omega\pi^\pm$ for the $\psipto\piz2(\pip\pim)$
candidate events. Resonance parameters are fixed to
their world averaged values~\cite{PDG}.}
\end{figure}

Figure~\ref{m5piall} (b) shows the  $\pp\piz$ invariant mass
distribution for
events satisfying $|m_{\gamma\gamma}-m_{\pi^o}|<0.03\gev/c^2$.
The fit is performed with an $\omega$ signal shape plus
a second order polynomial for the background,
and the number of $\omega$ signal events obtained is $386\pm23$.
The efficiency determined from MC simulation is 3.74\%
correcting for intermediate states, such as
$\omega f_2(1270)$ and $b_1^\pm\pi^\mp$, described
below.

Figure~\ref{m5piall} (c) shows the distribution of $\pip\pim$
invariant mass recoiling against the $\omega$, selected with the requirements
$|m_{\pip\pim\piz}-m_{\omega}|<0.05\gev/c^2$ and
$|m_{\omega\pi}-m_{b_1}|>0.2\gev/c^2$ to reject $b_1^\pm\pi^\mp$ events.
The invariant mass spectrum is fitted with a $\sigma$, a $f_2(1270)$
shape determined from MC simulation, and a second order polynomial to describe
other backgrounds. The number
of $f_2(1270)$ events obtained is $57\pm13$, and the detection efficiency
determined from MC simulation is 3.65\%.

Figure~\ref{m5piall} (d) shows the $\omega\pi^\pm$ invariant mass spectrum
with the requirement $|m_{\pip\pim\piz}-m_{\omega}|<0.05\gev/c^2$.
A clear $b_1^\pm$ signal is seen.
Fitting with a $b_1$ signal shape with the mass and width fixed to PDG values
plus a background polynomial,
the number of $b_1^\pm$ signal events is $202\pm21$, and
the detection efficiency determined from MC simulation is 3.24\%.
The branching fractions of these processes are determined to be:
\begin{eqnarray}
\BR(\psipto\piz 2(\pip\pim)) = (24.9\pm0.7)\times 10^{-4},\nonumber \\
\BR(\psipto\omega\pip\pim) = (8.4\pm0.5)\times 10^{-4},\nonumber \\
\BR(\psipto\omega f_2(1270)) = (2.3\pm0.5)\times 10^{-4},\nonumber \\
\BR(\psipto b_1^\pm\pi^\mp) = (5.1\pm0.6)\times 10^{-4}, \nonumber
\end{eqnarray}
where the errors are statistical.

\subsubsection{Signal analysis}
\begin{figure}[htbp] \centering
\includegraphics[width=0.45\textwidth]{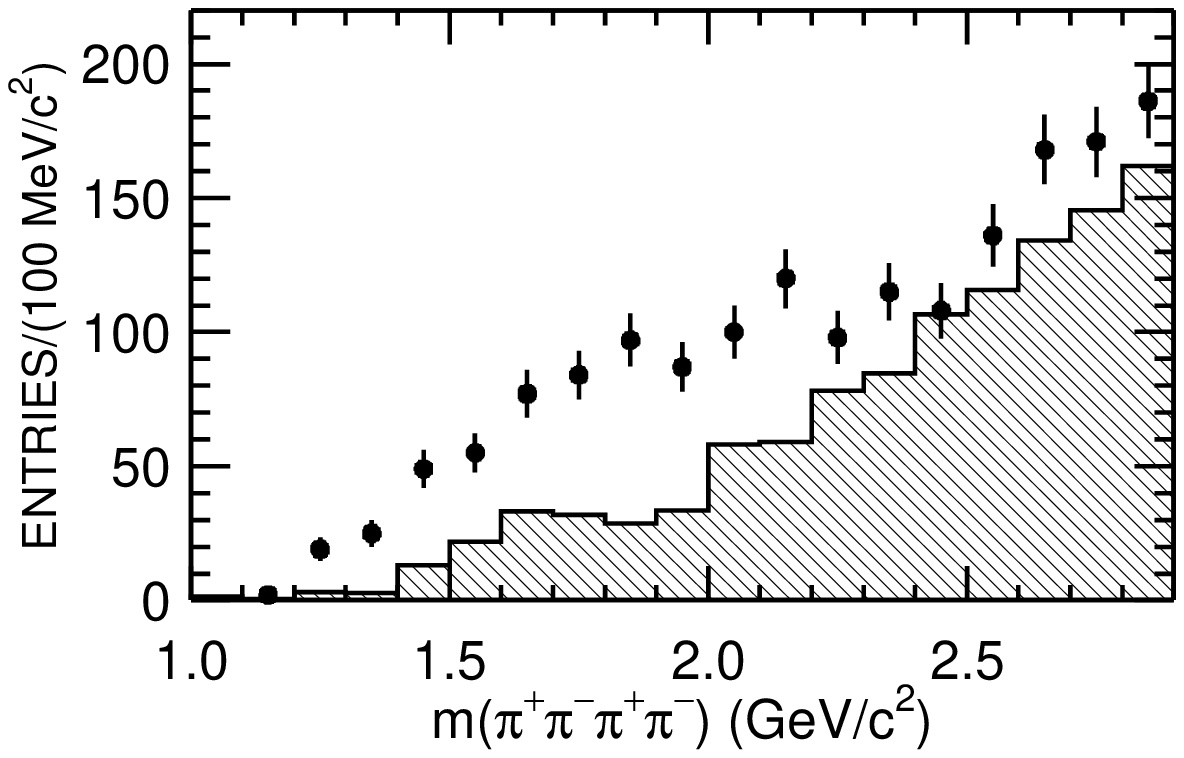}
\caption{\label{m4pi}The $2(\pp)$ invariant mass distribution for
  $\psipto\gamma2(\pp)$ candidate events (dots with error bars).  The
  shaded histogram includes contributions from the continuum
  (estimated from the data sample at $\sqrt{s}=3.65\gev$),
  $\psipto\piz 2(\pp)$, backgrounds remaining from
  $\pp\jpsi,\jpsi\ra\rhopi$ and $\piz\kskp$, and other unknown
  backgrounds estimated from the inclusive $\psi(2S)$ MC sample.}
\end{figure}

For $\psipto\gamma2(\pip\pim)$, candidate events require four charged
tracks, and each track must be identified as a pion.  The background
from $\psipto\ppjpsi$ is rejected by requiring
$|m_{recoil}^{\pi^+\pi^-} - m_{J/\psi}| > 0.05$ GeV/$c^2$.  After
selection, 1697 candidates remain, and the $2(\pp)$ invariant mass
distribution for the candidate events is shown in Fig.~\ref{m4pi}.
The backgrounds include contributions from the continuum (estimated
from the data sample at $\sqrt{s}=3.65\gev$), $\psipto\piz 2(\pp)$,
backgrounds remaining from $\pp\jpsi,\jpsi\ra\rhopi$ and $\piz\kskp$,
and the other unknown backgrounds assuming that they have the same
shape as that obtained from the inclusive $\psi(2S)$ MC sample.

Using the $\chi^2$ fitting method of Section~\ref{bkgs}, the number of
signal events is $583\pm41$.
The detection efficiency determined from MC simulation is 10.4\%, and
the branching fraction for this process is  determined to be:
\begin{equation}
\BR(\psipto\gamma2\pip\pim) = (39.6\pm2.8)\times 10^{-5}, \nonumber
\end{equation}
where the error is statistical.

\subsection{\boldmath $\psipto\gamma\kskp$}
For $\psipto\gamma\ks K^\pm\pi^\mp$, the main background comes from
$\psipto\ks K^\pm\pi^\mp\piz$, so we first measure $\psipto \ks
K^\pm\pi^\mp\piz$ in order to estimate its contamination to
$\psipto\gamma \ks K^\pm\pi^\mp$.

\subsubsection{Background estimation}
\begin{figure}[htbp]\centering
\includegraphics[width=0.45\textwidth]{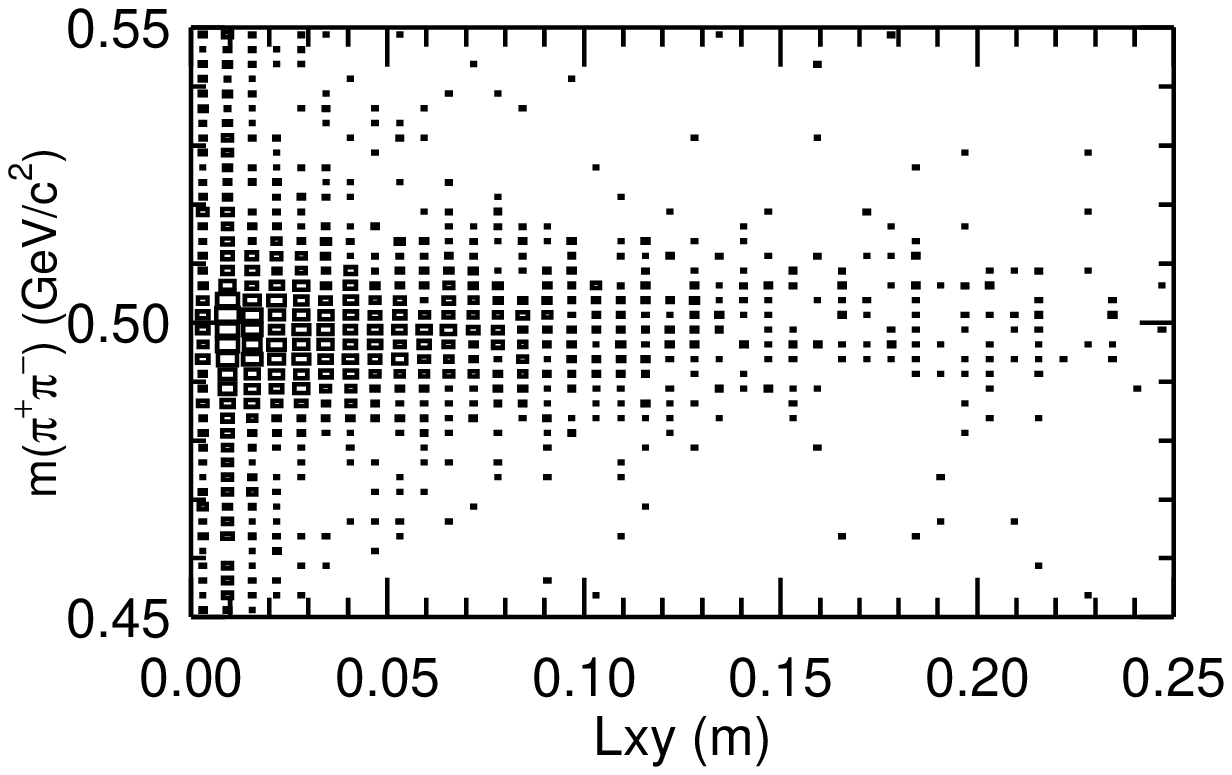}
\caption{\label{mks-lxy}The scatter plot of $\pp$ invariant mass versus the $\ks$ decay length for $\psipto\gamma\gamma\kskp$ candidate events.}
\end{figure}

For $\psipto\piz\kskp$, candidate events require four charged tracks
and two good photons.
Figure~\ref{mks-lxy} shows the scatter plot of
$\pi^+\pi^-$ invariant mass versus the decay length in the transverse
plane ($L_{xy}$) of $\ks$ candidates, where a clear $\ks$ signal is
observed.  Candidate events are required to have only one $\ks$
candidate satisfying the requirements $|m_{\pip\pim}-m_{K_S^0}|<0.015\gev/c^2$
and $L_{xy}>0.5$ cm.  After $\ks$ selection, the
remaining two tracks are identified using their
$\chi^2_{K\pi}$ values, \textit{i.e.}, if
$\chi^2_{\kap\pim}<\chi^2_{\pip\kam}$, the final state is considered
to be $\gamma\gamma\ks\kap\pim$; if $\chi^2_{\kam\pip}<\chi^2_{\pim\kap}$,
the final state is considered to be $\gamma \ks\kam\pip$, where
$\chi^2_{K\pi}=\chi^2_{PID}(K)+\chi^2_{PID}(\pi)$.

The confidence level of the 4C-fit must be greater than 1\%, and the
sum of the momentum of any $\pip$ and $\pim$ pair greater than
$550\mev/c$ to reject contamination from $\psipto\ppjpsi$ events.

\begin{figure}[htbp]\centering
\includegraphics[width=0.52\textwidth]{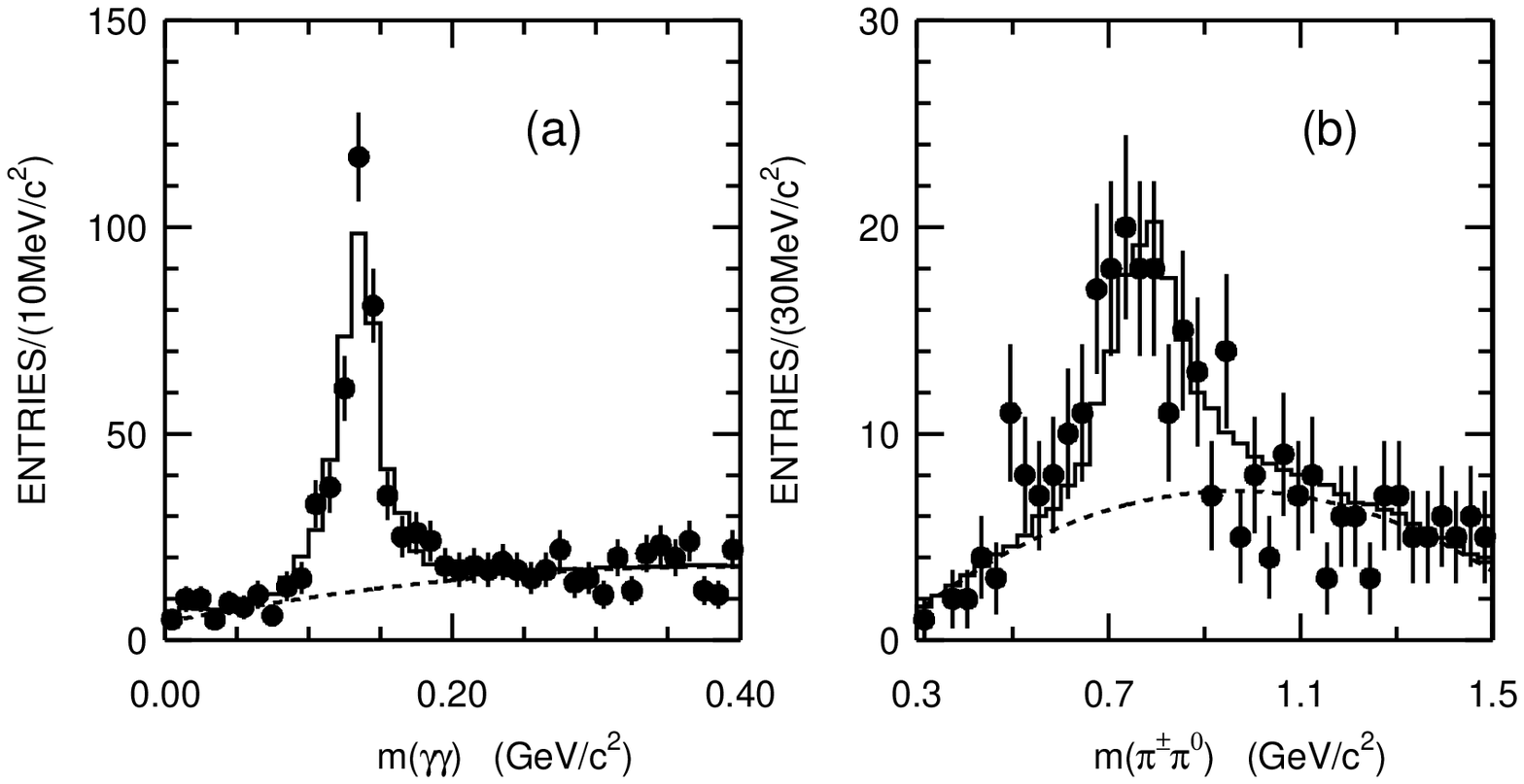}
\caption{\label{ggkskp}Invariant mass spectra of (a) $\gamma\gamma$
and (b) $\pi^\pm\piz$ for $\psipto\gamma\gamma\ks K^\pm\pi^\mp$
candidate events. Dots with error bars are data, the histograms are
the fits using signal shapes determined from Monte Carlo simulation
and second order polynomials for background, and the dashed curves are
the background shapes from the fit.}
\end{figure}

After requiring $|m_{\pip\pim}-m_{K_S^0}|<0.015\gev/c^2$, the
$\gamma\gamma$ invariant mass is shown in Fig.~\ref{ggkskp} (a), and a
clear $\piz$ signal is seen.  After requiring
$|m_{\gamma\gamma}-m_{\pi^o}|<0.03 \gev/c^2$, the $\pi^\pm\piz$ invariant
mass is shown in Fig.~\ref{ggkskp} (b), where there is a clear $\rho^\pm$
signal.

The $\gamma\gamma$ invariant mass distribution is fitted with a $\piz$
signal shape determined with MC simulation plus a second order
polynomial for the background, and the result is shown in
Fig.~\ref{ggkskp} (a).
The number of $\piz$ signal events fitted is $361\pm25$, and the
efficiency determined from MC simulation is 4.40\%, including the
effect of the intermediate $K^\pm\rho^\mp\ks$ state.

The $\pi^\pm\piz$ invariant mass distribution is fitted with a
$\rho^\pm$ signal shape determined with MC simulation plus a second
order polynomial for the background, and the fit result is shown in
Fig.~\ref{ggkskp} (b).  The number of $\rho^\pm$ signal events is
$100\pm20$, and the detection efficiency is 3.80\% determined from MC
simulation.  The branching fractions of these two processes are
determined to be
\begin{eqnarray}
\BR(\psipto\piz\ks K^\pm\pi^\mp) = (8.9\pm0.6)\times10^{-4},\nonumber \\
\BR(\psipto K^\pm\rho^\mp\ks) = (2.9\pm0.6)\times10^{-4}, \nonumber
\end{eqnarray}
where the errors are statistical.
\subsubsection{Signal analysis}
\begin{figure}[htbp]\centering
\includegraphics[width=0.45\textwidth]{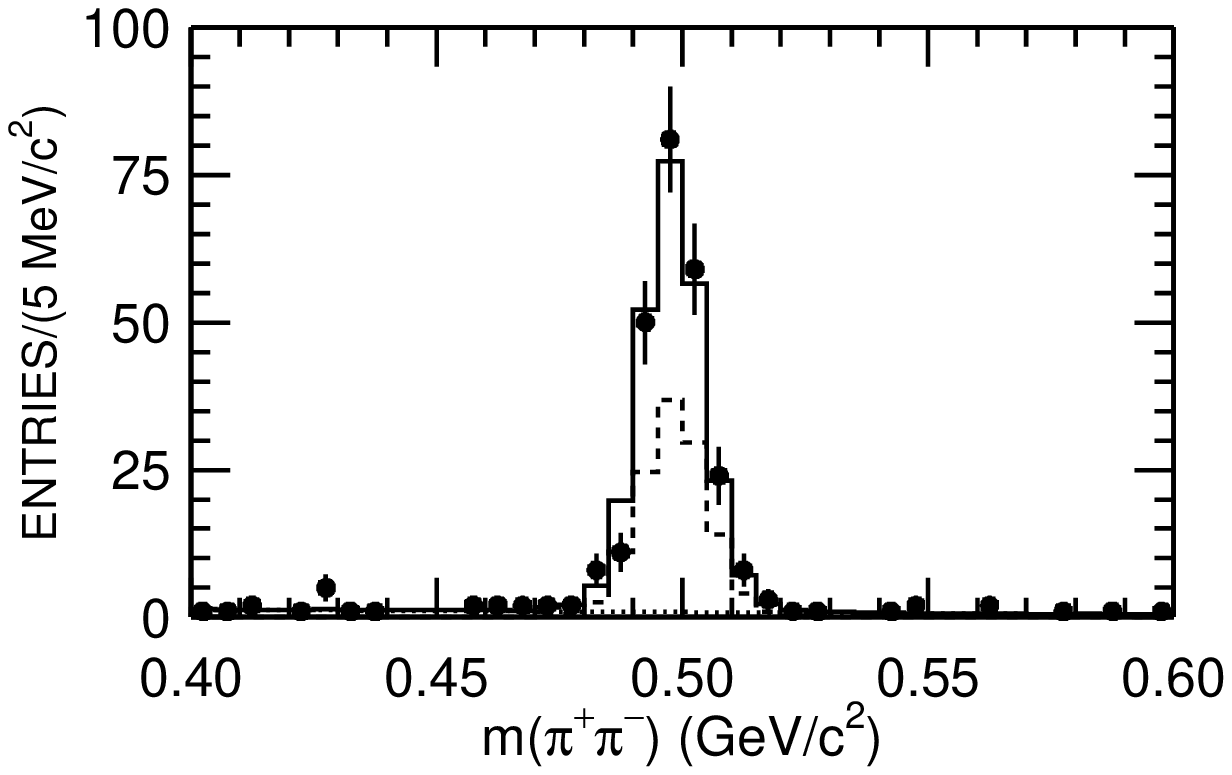}
\caption{The $\pip\pim$ invariant mass distribution for
$\psipto\gamma\kskp$ candidate events.  Dots with error bars are
data. The histogram is the fit with a histogram describing the $\ks$
shape obtained from MC simulation, the normalized histogram for
$\psipto\piz\kskp$ background (dashed histogram), and a Legendre
polynomial for the other smooth backgrounds (dotted histogram).}
\label{mksfit-gkskp}
\end{figure}

\begin{figure}[htbp] \centering
\includegraphics[width=0.45\textwidth]{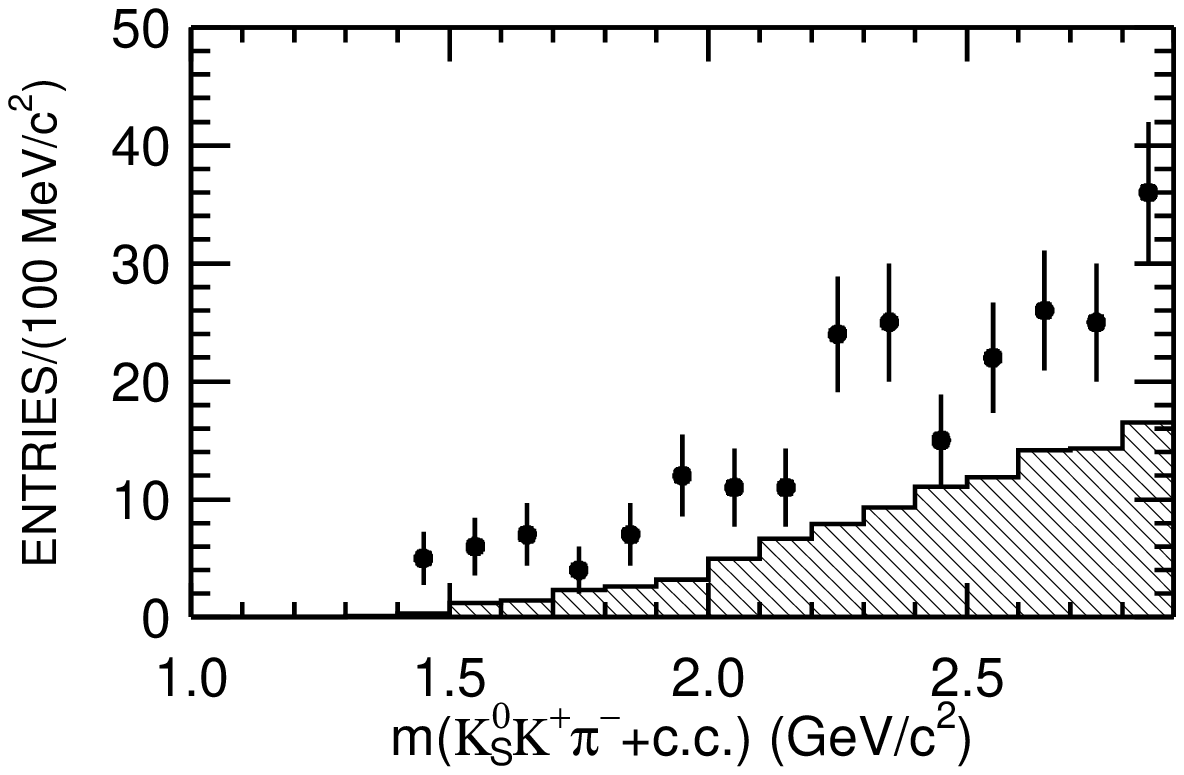}
\caption{\label{mkskp} The $\ks K^\pm\pi^\mp$ invariant mass
  distribution for $\psipto\gamma\kskp$ candidates (dots with error
  bars). The shaded histogram is the sum of backgrounds including
  $\psipto\piz\kskp$ and continuum background.}
\end{figure}

The event selection is similar to $\psipto\piz\kskp$, but only one
photon is required.  After event selection, the $\pip\pim$ invariant
mass distribution is shown in Fig.~\ref{mksfit-gkskp}.  A fit is
performed with a histogram describing the $\ks$ shape obtained from MC
simulation, the normalized histogram for $\psipto\piz\kskp$
background, and a Legendre polynomial for the other smooth
backgrounds. The fit yields $115\pm16$ events. The detection
efficiency is 4.83\%, and the branching fraction is determined to be:
$$\BR(\psipto\gamma\kskp)= (25.6\pm3.6)\times 10^{-5},$$
where the error is statistical.
Figure~\ref{mkskp} shows the $\ks K^\pm\pi^\mp$ invariant mass distribution
after event selection.

\subsection{\boldmath $\psipto\gamma\kap\kam\pip\pim$}
For $\psipto\gamma\kap\kam\pip\pim$, candidate events require four
charged tracks, among which two tracks must be identified as pions and
the other two tracks identified as kaons.  The background from
$\psipto\ppjpsi$ is rejected by requiring $|m_{recoil}^{\pi^+\pi^-} -
m_{J/\psi}| > 0.05$ GeV/$c^2$, the background from $\psi(2S) \to
\gamma2(\pip\pim)$ is rejected by requiring $\chi^2_{\gamma K^+
K^-\pi^+\pi^-} < \chi^2_{\gamma 2(\pi^+\pi^-)}$, and the background
from $\psi(2S) \to \gamma\kskp$ is rejected by requiring
$|m_{\pi^+\pi^-} - m_{K_S^0}| > 0.04$ GeV/$c^2$.

\begin{figure}[htbp] \centering
\includegraphics[width=0.45\textwidth]{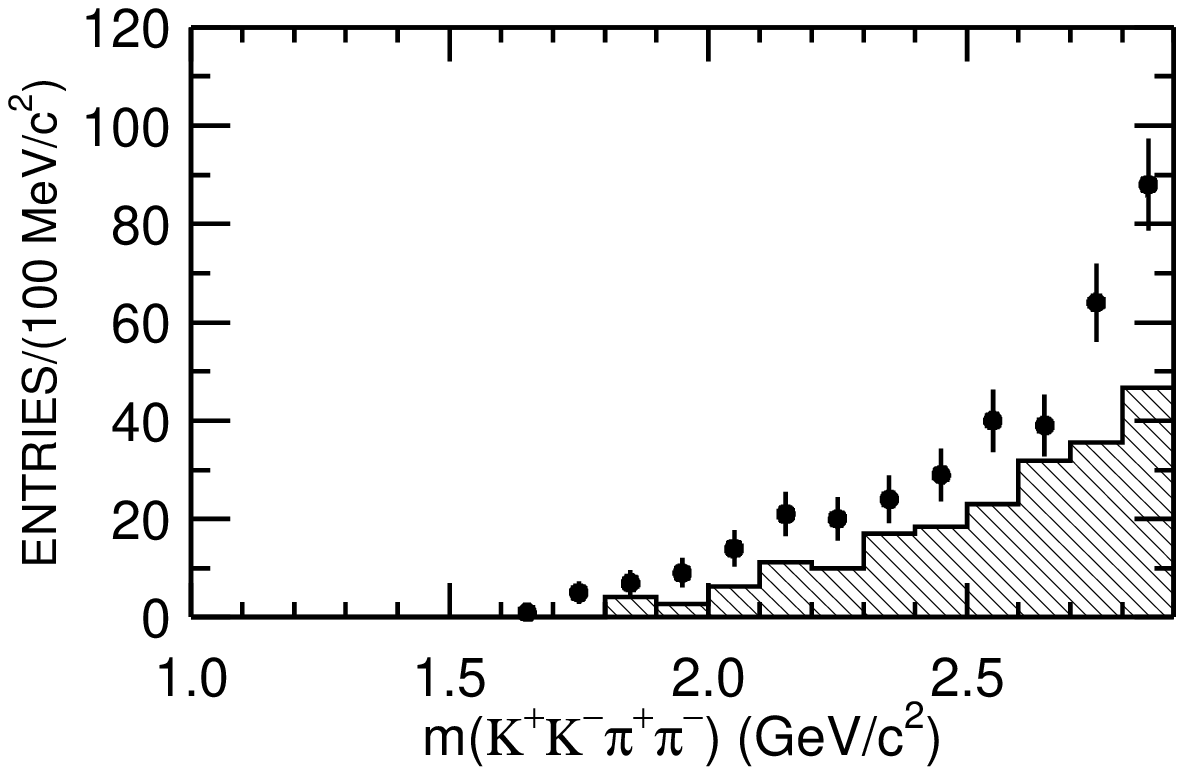}
\caption{\label{m2k2pi}The $\kap\kam\pp$ invariant mass distribution
  for $\psipto\gamma\kap\kam\pp$ candidates (dots with error bars). The
  shaded histogram is background which mainly comes from
  $\psipto\piz\kap\kam\pip\pim$.}
\end{figure}

Figure~\ref{m2k2pi} shows the $\kap\kam\pip\pim$ invariant mass distribution
after event selection, where 361 events are observed. The backgrounds
mainly come from $\psipto\piz\kap\kam\pip\pim$ final states including
intermediate states.

Using the $\chi^2$ fitting method of Section~\ref{bkgs}, the number of
signal events is $132\pm19$.
The detection efficiency determined from MC simulation is 4.94\%, and
the branching fraction for this process is  determined to be:
\begin{equation}
\BR(\psipto\gamma K^+ K^- \pip\pim ) = (19.1\pm2.7)\times 10^{-5},
\nonumber
\end{equation}
where the error is statistical.

\subsection{\boldmath $\psipto\gamma\kstarz\kap\pim+c.c.$ and $\psipto\gamma\kstarz\bar{\kstarz}+c.c.$ }
We apply the same event selection criteria as for
$\psipto\gamma\kap\kam\pip\pim$. For this decay channel, the main
background channels are from $\psipto\kstarz\kam\pip\piz +c.c.$ (phase
space), $\psipto\kstarz\kam\rho^+ +c.c.$, and $\psipto \kstarz
\kstarzb \piz$.

\begin{figure}[hbtp]\centering
\includegraphics[width=0.45\textwidth]{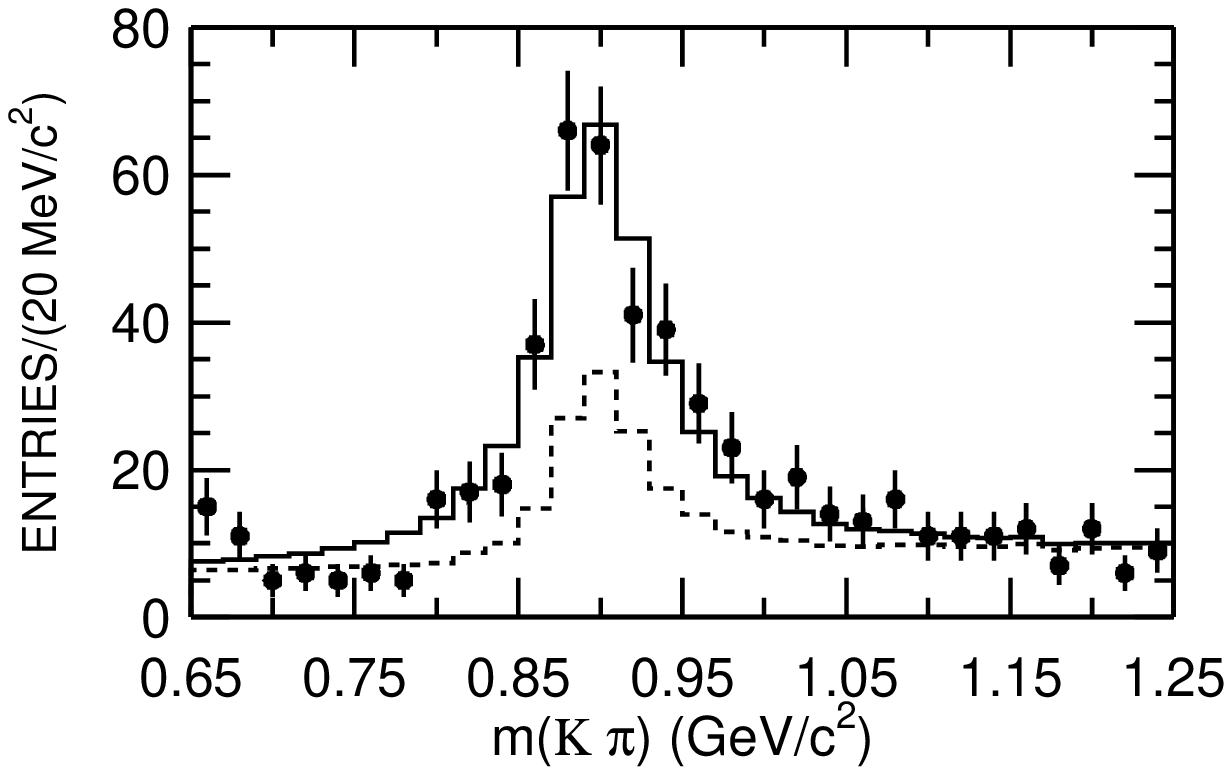}
\caption{\label{kstar-fit} The $K^\pm\pi^\mp$ invariant mass
distribution for $\psipto\gamma\kap\kam\pip\pim$ candidates. Dots with
error bars are data. The blank histogram is the result of a fit using
a Breit-Wigner function to describe the signal along with the sum of
normalized histograms from the $\psipto \kstarz\kam\pip\piz +c.c.$
(phase space), $\psipto \kstarz\kam\rho^+ +c.c.$, and $\psipto \kstarz
\kstarzb \piz$ background channels, and a second order Legendre
polynomial to describe other remaining backgrounds. The dashed
histogram is the fitted sum of all backgrounds.}
\end{figure}

Using a Breit-Wigner function to describe the signal along with the
sum of normalized histograms from the $\psipto \kstarz\kam\pip\piz
+c.c.$ (phase space), $\psipto \kstarz\kam\rho^+ +c.c.$, and
$\psipto \kstarz \kstarzb \piz$ background channels, and a second
order Legendre polynomial to describe other remaining backgrounds to
fit the $K^{\pm} \pi^{\mp}$ invariant mass spectrum, $237\pm39$
$\psipto\gamma\kstarz\kam\pip +c.c.$ candidate events are obtained,
as shown in Fig.~\ref{kstar-fit}. Events from the intermediate state
$\psipto\gamma\kstarz\kstarzb$ are counted twice, so an efficiency
correction must be made for this.

\begin{figure}[hbtp]\centering
\includegraphics[width=0.45\textwidth]{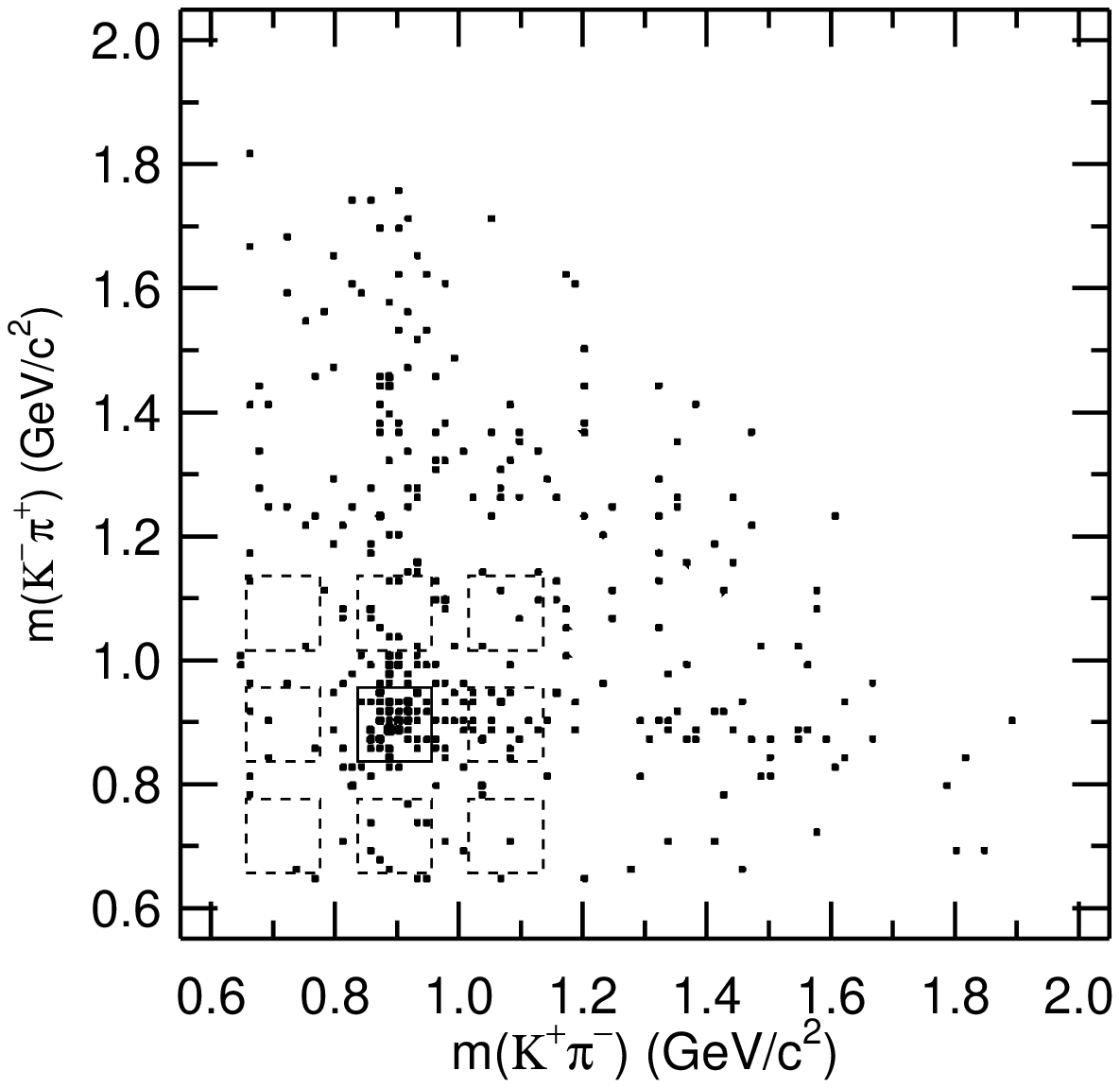}
\caption{\label{sandian} The scatter plot of $\kap\pim$ versus
$\kam\pip$ invariant mass of $\psipto\gamma\kap\kam\pip\pim$ candidate
events. The center box indicates the signal region for
$\psipto\gamma\kstarz\kstarzb$ events, and the other boxes are used for
background determination.}
\end{figure}

The scatter plot of $\kap\pim$ versus $\kam\pip$ invariant mass is
shown in Fig.~\ref{sandian}, where clear $\kstarz$ and $\kstarzb$
signals are seen.  The numbers of $\psipto\gamma\kstarz\kstarzb$ events
and background events are estimated from
the scatter plot.  The signal region is shown as a square box at
(0.896, 0.896)$\gev/c^2$ with a width of $60\mev/c^2$. Backgrounds are
estimated from sideband boxes, which are taken 60 MeV$/c^2$ away from
the signal box. Background in the horizontal or vertical sideband
boxes is twice that in the signal region.  If we subtract half the number
of events in the horizontal and vertical sideband boxes, we double
count the phase space background. Therefore the background is one-half
the number of events in the horizontal and vertical boxes plus one
fourth the number of the events in the four corner boxes.  After
subtraction, $41\pm8$ $\psipto\gamma\kstarz\kstarzb$ candidates are
obtained, and the efficiency is
$(2.75\pm0.06)\%$.  In simulating signal channels containing $\kstar$,
the shape of $\kstar$ is described by a P-wave relativistic
Breit-Wigner, with a width
$$\Gamma=\Gamma_0 \frac{m_0}{m} \frac{1+r^{2}p_0^2}{1+r^{2}
p^2}\Big[\frac{p}{p_0}\Big]^3,$$ where $m$ is the mass of the $K\pi$
system, $p$ is the momentum of the kaon in the $K\pi$ system,
$\Gamma_0$ is the width of the resonance, $m_0$ is the mass of the
resonance, $p_0$ is the momentum evaluated at the resonance mass, $r$
is the interaction radius, and $\frac{1+r^{2}p_0^2}{1+r^{2}p^2}$
represents the contribution of the barrier factor. The value
$r=(3.4\pm0.6\pm0.3) (\gev/c)^{-1}$ measured by the $\kam\pip$
scattering experiment \cite{r-aston} is used as an approximate
estimation of the interaction radius $r$.

Taking into consideration the effect of the intermediate channel,
$\psi(2S) \to \gamma K^{*0}\bar{K^{*0}}$,
the efficiency for $\psipto\gamma\kstar\kam\pip +c.c.$ is 6.86\%, and
we obtain
the branching fractions:
\begin{eqnarray}
\BR(\psipto\gamma\kstar\kam\pip +c.c.) =  (37.0\pm6.1)\times
10^{-5}\nonumber \\
\BR(\psipto\gamma\kstarz\kstarzb) =  (24.0\pm4.5)\times 10^{-5},\nonumber
\end{eqnarray}
where the errors are statistical.

\subsection{\boldmath $\psipto\gamma2(\kap\kam)$}
\begin{figure}[htbp] \centering
\includegraphics[width=0.45\textwidth]{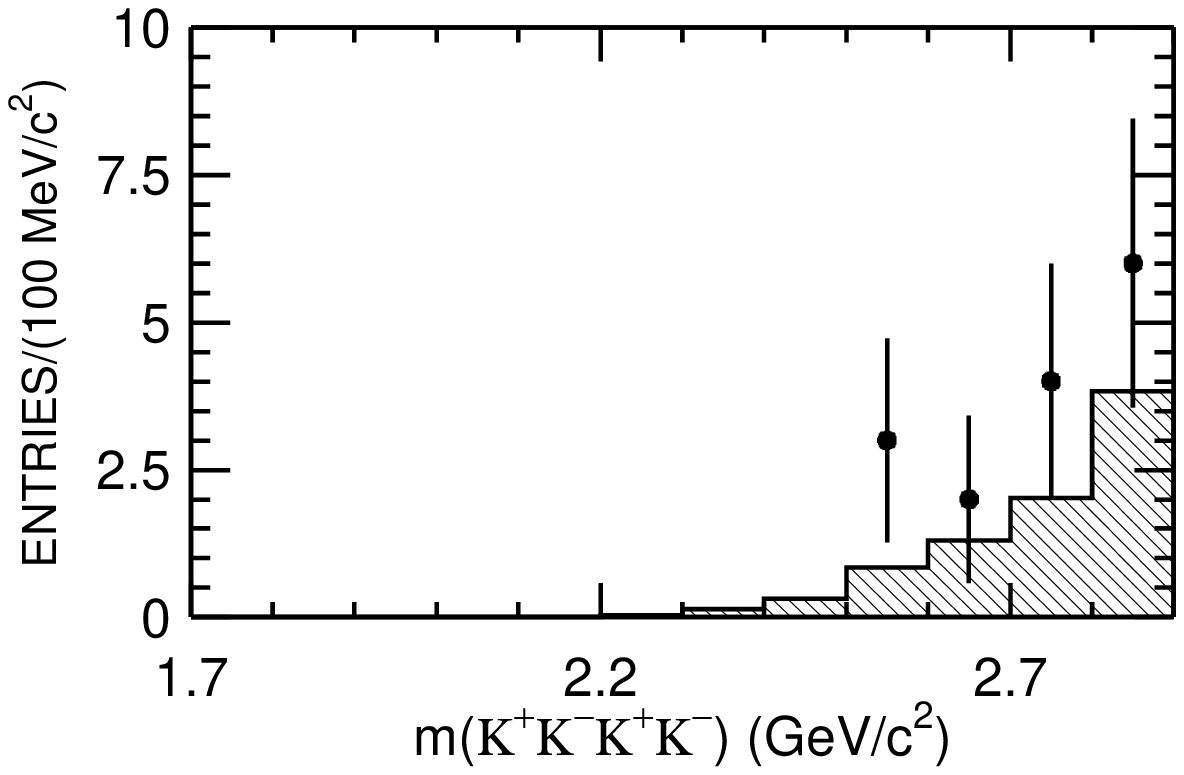}
\caption{\label{m4k}Invariant mass distribution of $2(\kap\kam)$ for
$\psipto\gamma2(\kap\kam)$ candidates (dots with error bars). The
shaded histograms is background mainly from $\psi(2S) \to
\pi^02(K^+K^-)$.}
\end{figure}

For $\psipto\gamma2(\kap\kam)$, the candidate events must have four
charged tracks, and every track must be identified as a kaon. The
backgrounds from $\psipto\gamma2(\pip\pim), \gamma\pip\pim\kap\kam$
are rejected by requiring the $\chi^2_{4C}$ for the signal channel
to be less than those for backgrounds. There are 15 events observed
after event selection, and the $2(\kap\kam)$ invariant mass
distribution is shown in Fig.~\ref{m4k}. The detection efficiency
for this channel is 2.93\%.

The dominant background comes from $\psipto\piz 2(\kap\kam)$.  Using
the branching fraction measured by CLEO~\cite{chic23hs}, the
estimated number of background events remaining is $8\pm2$. To
measure the continuum contribution in this channel, the continuum
data at $E_{cm}=3.65\gev$ is analyzed using the same criteria as for
$\psip$ data, and no events survive.  It is also found that no
events survive from the simulated 14 million inclusive $\psip$ decay
MC sample.  The upper limit on the number of
$\psipto\gamma2(\kap\kam)$ events is 14 at the 90\% C.L., and the
corresponding upper limit on the branching fraction after
considering systematic uncertainties is
\begin{equation}
\BR(\psipto\gamma2(\kap\kam))<4.0\times 10^{-5}. \nonumber
\end{equation}

\subsection{\boldmath $\psipto\gamma\pip\pim\ppb$}
For $\psipto\gamma\pip\pim\ppb$, there must be four good charged
tracks, and two of them must be identified as a proton anti-proton
pair.  The backgrounds from $\gamma2(\pip\pim)$ and
$\gamma\pip\pim\kap\kam$ are rejected by requiring $\chi^2_{4C}$ for
the signal channel to be less than for the background channels.  To
eliminate possible contamination from $\psipto\pi^+\pi^- J/\psi,J/\psi
\to\gamma\ppb$, we require
$|m^{\pp}_{recoil}-m_{J/\psi}|>0.02\gev/c^2$.

\begin{figure}[htbp] \centering
\includegraphics[width=0.45\textwidth]{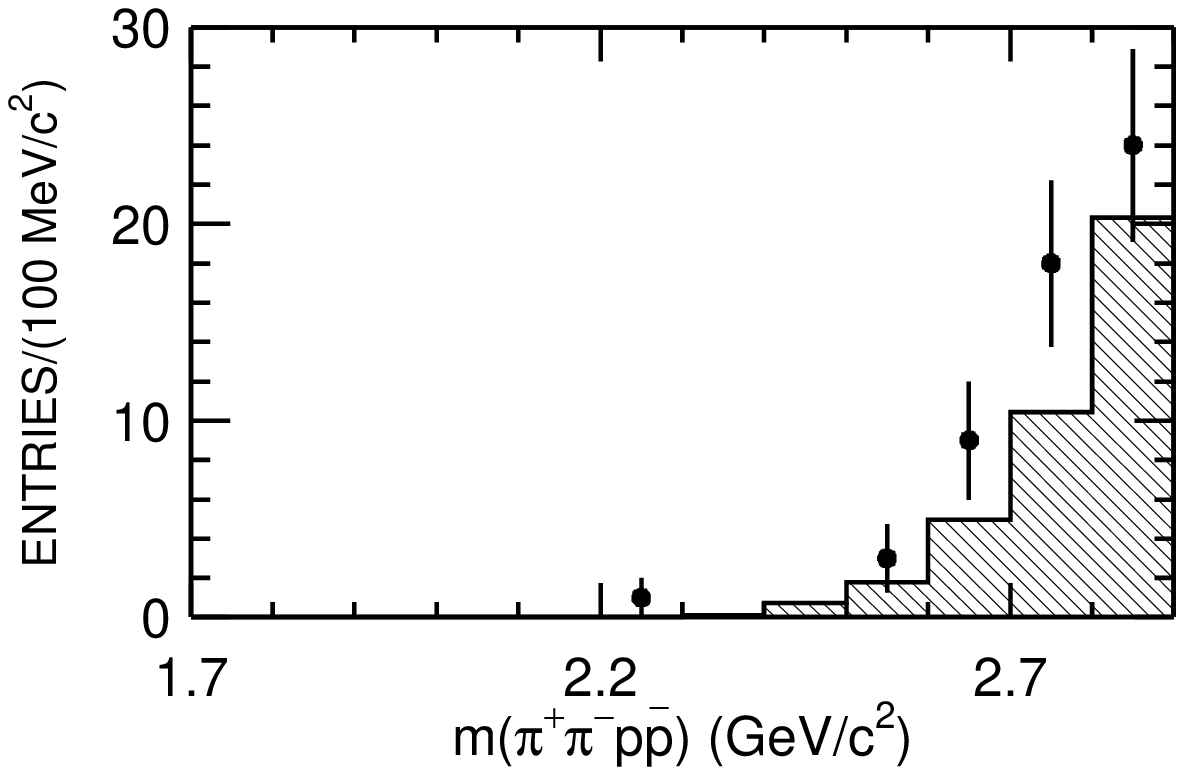}
\caption{\label{m2pippb}Invariant mass distribution of $\pp\ppb$ for
$\psipto\gamma\pp\ppb$ candidates (dots with error bars). The shaded
histogram is background mainly from $\psipto\piz\pip\pim\ppb$ and
$\psipto\ppjpsi,\jpsito\gamma\ppb$.}
\end{figure}

Figure~\ref{m2pippb} shows the $\pp\ppb$ invariant mass distribution
with 55 events
after event selection.
The detection efficiency for this channel is 4.47\%.

The dominant backgrounds are $\psipto\piz\pip\pim\ppb$ and background
remaining from
$\psipto\ppjpsi,\jpsito\gamma\ppb$. The detection efficiencies for
these two background channels are determined by MC simulation to be
0.35\% and 0.18\%, respectively. For the first background channel,
using $\BR(\psipto\piz\pip\pim\ppb)$ measured by CLEO~\cite{chic23hs},
the number of background events remaining is estimated to be $35.8\pm
3.5$. Similarly, the estimated number of background events from
$\psipto\ppjpsi,\jpsito\gamma\ppb$ is $1.7\pm0.4$.  Subtracting
backgrounds, the number of $\psipto\gamma\pip\pim\ppb$ events is
$17\pm 7$, and the corresponding branching fraction is
\begin{equation}
\BR(\psipto\gamma\pip\pim\ppb) = (2.8\pm 1.2)\times10^{-5}, \nonumber
\end{equation}
where the error is statistical.
\subsection{\boldmath $\psipto\gamma3(\pip\pim)$}
\begin{figure}[htbp] \centering
\includegraphics[width=0.45\textwidth]{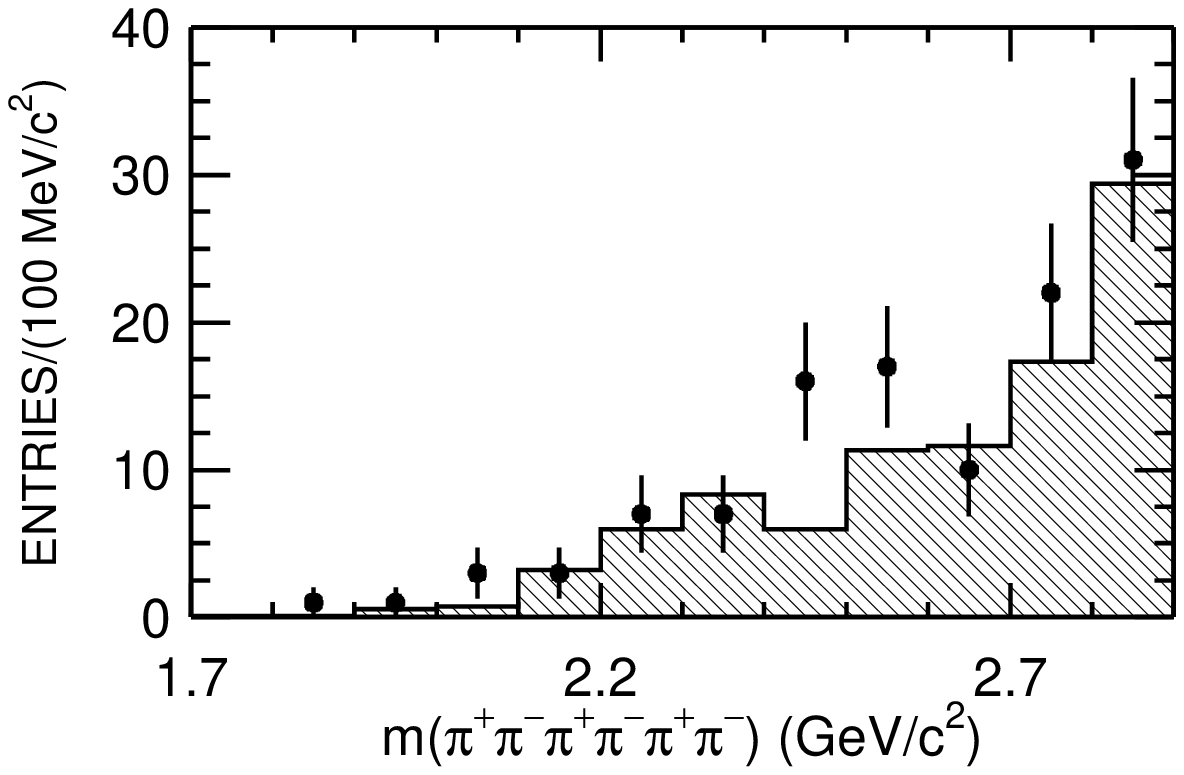}
\caption{\label{m6pi}The $3(\pp)$ invariant mass distribution for
$\psipto\gamma3(\pp)$ candidates (dots with error bars). The shaded
histogram is background mainly from processes with multi-photon final
states, such as $\psipto\piz 3(\pip\pim), \gamma\piz 3(\pip\pim),$ and
$\piz\piz 3(\pip\pim)$.}
\end{figure}

For $\psipto\gamma3(\pip\pim)$, six charged tracks are required. The
backgrounds from $\psipto\ppjpsi$ and $\psipto\ks\ks\pip\pim$ are
removed by eliminating events having the recoil mass of any pion
pair satisfying $|m^{\pip\pim}_{recoil}-m_{J/\psi}|<0.05\gev/c^2$ or
having a pion pair in the $K_S^0$ mass region from 0.47 to 0.53
GeV/$c^2$.  The remaining backgrounds mainly come from processes
with multi-photon final states, such as $\psipto\piz 3(\pip\pim),
\gamma\piz 3(\pip\pim),$ and $\piz\piz 3(\pip\pim)$. Their
contaminations are estimated using MC simulation.  Figure~\ref{m6pi}
shows the $3(\pp)$ invariant mass distribution after event selection
with 118 events observed.  The detection efficiency for this channel
is 1.97\%. Using the $\chi^2$ fitting method described in
Section~\ref{bkgs}, the upper limit on the number of signal events
is 45 at the 90 \% C.L., and the upper limit on the branching
fraction  after considering systematic uncertainties is $B(\psi(2S)
\to \gamma 3(\pi^+\pi^-)) < 17 \times 10^{-5}$.

\subsection{\boldmath $\psipto\gamma2(\pip\pim)\kap\kam$}
\subsubsection{Background estimation}
 First, background from $\psipto \eta \jpsi$,
$\eta \to \gamma \pi^+\pi^-$, $\jpsito \pi^+ \pi^- K^+ K^-$ is
rejected by requiring $m_{2(\pip\pim)\kap\kam}<2.9$ GeV$/c^2$. The
dominant backgrounds remaining are $\psipto\piz 2(\pip\pim)\kap\kam$
and $\gamma\piz 2(\pip\pim)\kap\kam$. Branching fractions for these
are not currently available, so they are measured using our $\psip$
data sample.

For $\psipto\piz 2(\pip\pim)\kap\kam$, the number of good photons is
required to be $N_\gamma=2$ or $3$. A kinematic fit is performed
under the $\psip\to\gamma\gamma2(\pip\pim)\kap\kam$ hypothesis
running over all selected photons, and the combination with the
smallest $\chi^2$ is retained. Background from $\psipto \pi^+ \pi^-
\jpsi$, $\jpsito \gamma \gamma 2(\pi^+ \pi^-) K^+K^-$ is rejected by
requiring $|m^{\pi^+\pi^-}_{recoil}-m_{\jpsi}|>0.05$
$\hbox{GeV}/c^2$. The possible backgrounds from
$\psip\to\gamma2(\pip\pim)\kap\kam$ and $3\gamma2(\pip\pim)\kap\kam$
are rejected by requiring the $\chi^2$ value for the signal to be
less than those for the backgrounds.  To remove backgrounds from
$\psip\to\gamma\chi_{cJ}$ decays, we require
$m_{2(\pip\pim)\kap\kam}<3.38$ GeV$/c^2$.

  Eight main peaking background channels, including $\psipto \eta
\jpsi$, $\pi^0 \jpsi$ and $\gamma \chi_{cJ}$ to decay into the same
final states, are simulated and fitted using the same procedure and
selection criteria, and $9.8\pm4.5$ background events are obtained.
Using the 14 million inclusive MC sample, $14.7\pm5.7$ background
events are found, which is consistent with the simulation result
within the statistical error.

\begin{figure}[htpb]\centering
\includegraphics[width=0.45\textwidth]{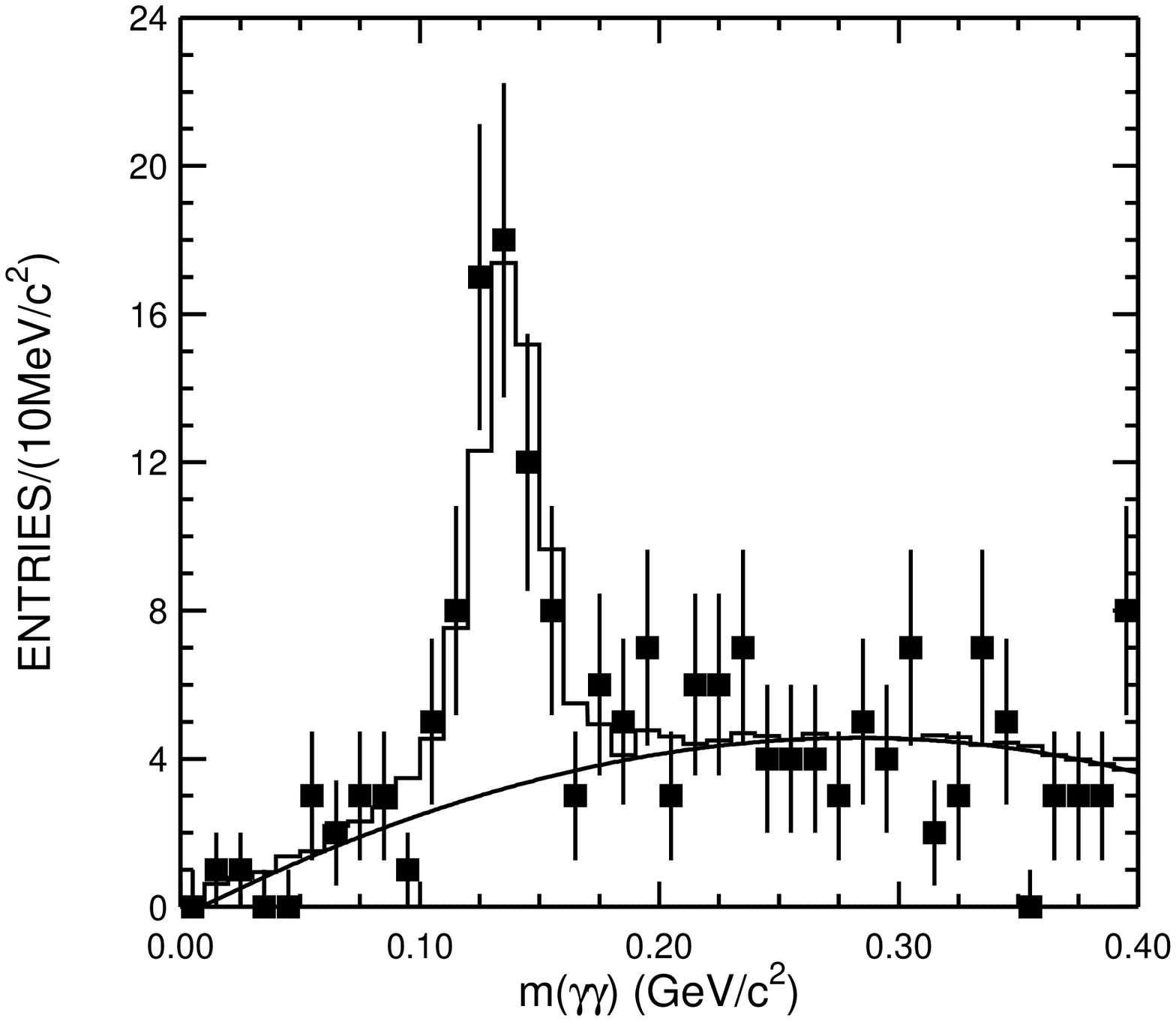}
\caption{\label{5pi2k} The $\gamma\gamma$ invariant mass
distribution for $\psipto\gamma\gamma2(\pip\pim)\kap\kam$ candidate
events. Dots with error bars are data, and the blank histogram is
the fit with a $\piz$ signal shape determined from MC simulation
plus a third order polynomial for the background. The curve is the
fitted background.}
\end{figure}

 Figure~\ref{5pi2k} shows the
$\gamma\gamma$ invariant mass distribution for $\psipto\piz
2(\pip\pim)\kap\kam$ candidate events. A fit is performed with a
$\piz$ signal shape determined from MC simulation plus a third order
polynomial for the background, and the number of $\piz$ signal
events is determined to be $57.4\pm9.8$. After subtracting peaking
backgrounds, the number of signal events is $47.6\pm10.8$.

The detection efficiency determined from MC simulation using a phase
space generator is $0.41\%$ below $m_{2(\pi^+\pi^-)K^+K^-}=3.38
\textrm{ GeV}/c^2$, and the branching fraction is determined to be:
\begin{equation}
\BR(\psipto\piz2(\pip\pim)\kap\kam) = (8.39\pm1.91)\times 10^{-4}
\nonumber
\end{equation}
with the requirement $m_{2(\pi^+\pi^-)K^+K^-}<3.38 \textrm{
GeV}/c^2$, where the error is statistical. The effect of possible
intermediate resonances is not considered. Assuming phase space
production, the branching fraction extrapolated to the full
$m_{2(\pip\pim)\kap\kam}$ energy region is determined to be
$\BR(\psipto\piz2(\pip\pim)\kap\kam) = (11.5\pm2.6)\times 10^{-4}$.

For $\psipto\gamma\piz2(\pip\pim)\kap\kam$, the number of good
photons is required to be $N_\gamma=3$ or 4. A kinematic fit is
performed under the $\psip\to3\gamma2(\pip\pim)\kap\kam$ hypothesis
running over the selected photons; the combination with the smallest
$\chi^2$ is retained. Possible backgrounds from $\psipto
(n\gamma)2(\pip\pim)\kap\kam$ with $n=1$  $2$ and $4$ and from
$\psipto3\gamma 3(\pip\pim),~3\gamma\pip\pim2(\kap\kam)$, and
$3\gamma K^{\pm}\pi^{\mp}2(\pip\pim)$ are rejected by requiring that
the $\chi^2$ of the signal is less than those of the backgrounds.
Background from $\psip\to\pip\pim J/\psi,~\jpsi\to3\gamma \pip\pim
\kap\kam$ is rejected by requiring
$|m^{\pim\pip}_{recoil}-m_{\jpsi}|>0.05$ GeV/$c^2$, and backgrounds
from
$\psipto\piz\jpsi,~\eta\jpsi,~\piz\piz\jpsi,~\jpsi\to2(\pip\pim)\kap\kam$
are rejected with the requirement
$|m_{2(\pip\pim)\kap\kam}-m_{\jpsi}|>0.05$ GeV/$c^2$. We select the
$\pi^0$ from the $\gamma\gamma$ combinations as the one with
$m_{\gamma\gamma}$ invariant mass closest to $m_{\pi^0}$. To remove
backgrounds from $\psip\to\gamma\chi_{cJ}$ decays,
$m_{\piz2(\pip\pim)\kap\kam}<3.38$ GeV$/c^2$ is required.

After event selection, no significant $\piz$ candidates are
observed. A fit with a $\piz$ shape determined from MC simulation
plus a second order Legendre polynomial for background yields
$27.1\pm8.5$ events.  Fitting in the same way a histogram of 20 MC
simulated background modes, $21.2\pm10.6$ background events are
obtained. The detection efficiency determined from MC simulation is
$5.8\times 10^{-4}$. The upper limit on the branching fraction at
the $90\%$ C.L., determined using {\bf POLE} \cite{pole} and
including systematic uncertainties, is
$$B(\psipto\gamma\piz2(\pip\pim)\kap\kam)<3.1\times 10^{-3}.$$

\subsubsection{Signal analysis}
For $\psipto\gamma2(\pip\pim)\kap\kam$, six charged tracks are
required, and two of them must be identified as kaons.  The
background from $\psipto\ppjpsi,J/\psi \to \gamma +$\:four charged
particles is removed by requiring $|m^{\pi^+\pi^-}_{recoil} -
m_{J/\psi}| > 0.05$ GeV$/c^2$, and the backgrounds from $\psi(2S)
\to \gamma 3(\pip\pim), \gamma K^\pm\pi^\mp 2(\pip\pim)$ and
$\gamma2(\kap\kam) \pip\pim$ are rejected by requiring the $\chi^2$
values for the signal to be smaller than for the backgrounds. The
background from $\psi(2S)\to\eta J/\psi\to \gamma
2(\pip\pim)\kap\kam$ is rejected by requiring
$m_{2(\pip\pim)\kap\kam}<2.9\textrm{GeV}/c^2$.

\begin{figure}[htbp] \centering
\includegraphics[width=0.45\textwidth]{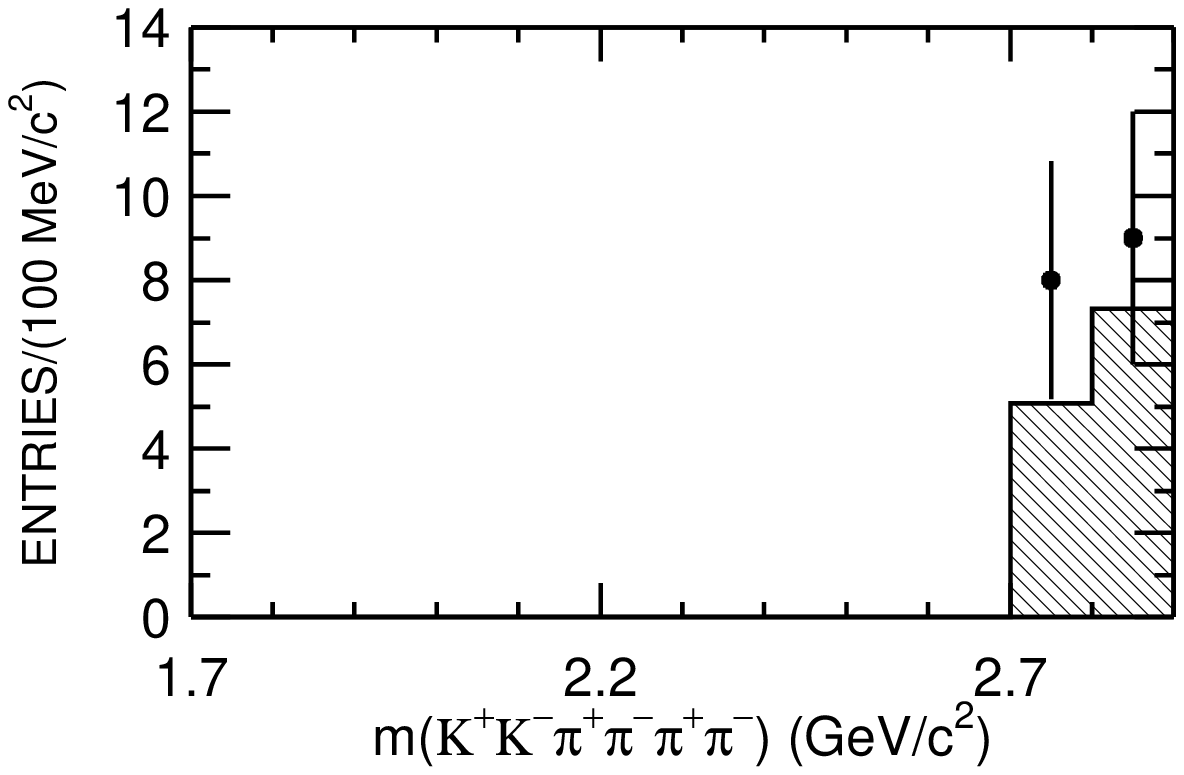}
\caption{\label{m2k4pi}The $2(\pp)\kap\kam$ invariant mass distribution
for $\psipto\gamma2(\pp)\kap\kam$ candidates (dots with error
bars). The shaded histogram is
background mainly from  $\psipto\piz
2(\pip\pim)\kap\kam$ and $\gamma\piz 2(\pip\pim)\kap\kam$.}
\end{figure}

Figure~\ref{m2k4pi} shows the $2(\pp)\kap\kam$ invariant mass
distribution, where the shaded histogram is background mainly from
$\psipto\piz 2(\pip\pim)\kap\kam$ and $\gamma\piz
2(\pip\pim)\kap\kam$. For the $\psipto\piz 2(\pip\pim)\kap\kam$
background channel,  the branching fraction of $(8.39\pm1.91)\times
10^{-4}$ is used  since the MC sample is produced with
$m_{2(\pip\pim)\kap\kam}<3.38$ $\hbox{GeV}/c^2$. After subtracting
all backgrounds, 17 events are obtained.
 The detection efficiency for this channel is 0.69\%. The upper
limit on the number of signal events is 15.5 at the 90\% C.L., and
the branching fraction after considering systematic uncertainties is
$B(\psi(2S) \to \gamma 2(\pi^+ \pi^-)K^+K^-) < 22 \times 10^{-5}$.

\begin{table*}
\begin{center}
\caption{\label{syserr}Summary of systematic errors (\%), where WR,
$\eff_\gamma$, $\ks$ rec., and MC denote the wire resolution, photon
efficiency, the error for $\ks$ reconstruction, and MC statistics,
respectively. The sixth column gives the uncertainties due to the
$\chi^2$ fits or the $\kstar$ fit.}
\begin{tabular}{lcccccccccr} \hline \hline
Mode  & WR & $\eff_\gamma$ & PID & $\ks$ rec. &  fit & Branching
Fractions &Background&$N_{\psi(2S)}$&MC&Total\\\hline
$\gamma \ppb$ &6.3&2.0&4.0&---&---&---&9.4&4.0&0.5&12.8 \\
$\gamma\ppb\piz$  &11.6 & 6.0 & 4.0 & ---& ---& ---&14.3 & 4.0 &3.0 & 20.4\\
$\gamma 2(\pp)$ &5.0&2.0&8.0&---&3.0 ($\chi^2$ fit)&---&6.4&4.0&1.0&12.7\\
$\gamma \ks\kap\pim +c.c.$  &5.0&2.0&---&3.4&---&---&11.8&4.0&1.0&14.1\\
$\gamma \kap\kam\pip\pim$ &10.7 &2.0  &8.0&--- & 3.0 ($\chi^2$ fit)&--- & 17.2 &4.0&2.2&22.6\\
$\gamma \kstarz\kap\pim +c.c.$&10.7 &2.0& 8.0&--- & 8.8 ($\kstar$ fit)&--- & 10.0&4.0&1.2&19.5\\
$\gamma \kstarz\overline{K}^{*0}$&10.7 &2.0& 8.0&--- & --- &---& 10.0&4.0&1.1&17.4\\
$\gamma\pp\ppb$ &10.4&2.0&8.0&---&---&---  &20.6&4.0&1.1&24.9\\
$\gamma2(\kap\kam)$  &11.1&2.0&8.0 &---&---&---&14.2&4.0&1.6&20.3\\
$\gamma3(\pp)$&5.0&2.0&---&---&---&---&---&4.0&2.2&7.0\\
$\gamma2(\pp)\kap\kam$&8.7&2.0&4.0&---&3.0 ($\chi^2$ fit)&--- &25.0&4.0&3.0&27.5 \\
$\gamma\piz 2(\pip\pim)\kap\kam$& 9.1&6.0&4.0&---&---&---&6.0&4.0&3.1&14.0\\
$\ppb\piz\piz$  &11.7 & 8.0 & 4.0 & ---& ---& ---& 4.3 & 4.0 & 1.7 & 15.9\\
$\piz 2(\pp)$ &10.0&4.0&8.0&---&---&---&1.8&4.0&1.0&14.2\\
$\omega\pip\pim$   &10.0&4.0&8.0&---&---& 0.8 ($\omega$ Br)&2.0&4.0&1.0&14.2\\
$\omega f_2(1270)$ &10.0&4.0&8.0&---&---&3.1 ($f_2$ Br)&10.0&4.0&1.0&17.5\\
$b_1^\pm\pi^\mp$ &10.0&4.0&8.0&---&---&0.8 ($b_1$ Br)&1.0&4.0&1.0&14.1\\
$\piz\ks K^\pm\pi^\mp$&10.0&4.0&---&3.4 &---&---&3.0&4.0&1.5&12.5\\
$K^\pm\rho^\mp\ks$&10.0&4.0&---&3.4&---&---&8.0&4.0&1.5&14.5\\
$\piz 2(\pip\pim)\kap\kam$&13.5&4.0&4.0&---&---&---&13.3&4.0&1.1&20.2\\
\hline \hline
\end {tabular}
\end{center}
\end{table*}

\section{Systematic errors}
Systematic errors on the branching fractions, listed in
Table~\ref{syserr}, mainly originate from
the MC statistics, the track error matrix, the kinematic fit,
particle identification, the photon efficiency, the $\chi^2$ fit method,
the uncertainty of the branching fractions of intermediate
states (taken from the PDG~\cite{PDG}), the uncertainty of the background
estimation, and the total number of $\psip$ events.

\begin{enumerate}
\item The systematic error caused by the MDC tracking and the
kinematic fit is estimated by using simulations with different MDC
wire resolutions~\cite{simbes}. The systematic error ranges from 5\%
to 13.5\% depending on the number of charged tracks in the different channels.

\item The photon detection efficiency was studied
with $J/\psi \to \pi^+\pi^-\pi^0$ events~\cite{simbes}, and
the difference between data and MC simulation is about $2\%$ for
each photon.

\item Pure $\pi$ and $K$ samples were selected, and the particle identification
efficiency was measured as a function of track momentum. On the
average, a $1.3\%$ efficiency difference per $\pi$ track and a
$1.0\%$ difference per $K$ track are observed between data and MC
simulation. We take $2.0\%$ for each charged particle identification
as a conservative estimate of the systematic error.

\item In order to estimate the systematic error caused by the
differences of the $\chi^2$ distributions between data and MC
simulation, we use selected samples of $\psip \to \gamma \chi_{c0}$,
$\chi_{c0}\to K^+ K^- \pi^+ \pi^-$ and $\pi^+ \pi^-\pi^+ \pi^-$ to
compare the $\chi^2$ shapes of data and MC, because these two samples
have similar final states and sufficient statistics.  The difference
is about 3\%, which is taken as the systematic error of the $\chi^2$
fit method. We also performed an input-output study of the $\chi^2$
fit, and found the difference between input and output values is
very small ($<0.5\%$) and is neglected.

\item The background uncertainties are estimated by changing
the order of the polynomial or the fitting range used. The
errors on the branching fractions of the main backgrounds ($\psi(2S)
\to \pi^0 + hs$) have also been considered and included. The
uncertainty of the background estimation varies from 1\%-25\%
depending on the channel and background level.

\item The uncertainty of the total number of $\psip$ events is
4\%~\cite{pspscan}.
\end{enumerate}

 Adding up all these sources in quadrature, the total systematic
errors range from 7\% to 28\% depending on the channel.

\section{Results and conclusions}
\begin{figure}\centering
\includegraphics[width=0.45\textwidth]{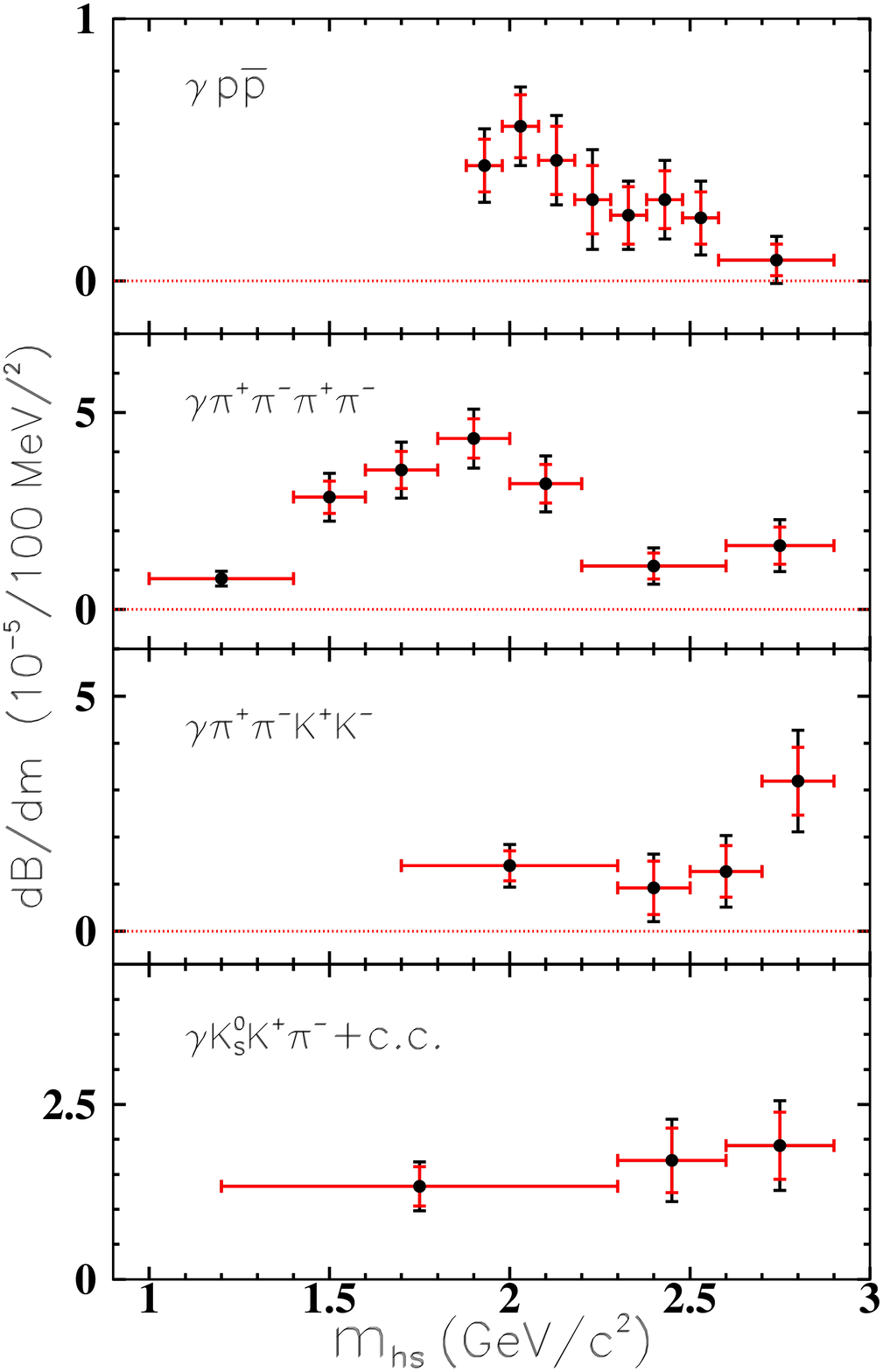}
\caption{ \label{diffbr} Differential branching fractions for $\psip$
decays into $\gamma\ppb$, $\gamma 2(\pip\pim)$, $\gamma\kap\kam\pip\pim$,
and $\gamma\kskp$ Here $m_{hs}$ is the invariant mass of the hadrons
in each final state. For each point, the smaller vertical error is
the statistical error, while the bigger one is the sum of statistical
and systematic errors. }
\end{figure}

Figure~\ref{diffbr} shows the differential branching fractions for
$\psip$ decays into $\gamma\ppb$, $\gamma 2(\pp)$,
$\gamma\kap\kam\pip\pim$, and $\gamma\kskp$, and the numbers of
events extracted for each decay mode with $m_{hs}<2.9\gev/c^2$ are
listed in Table~\ref{Tot-nev}.  Broad peaks, which are similar to
those observed in $\jpsi$ decays into the same final
states~\cite{jpsi-gppb,g4pi}, appear in the $m_{\ppb}$ and $m_{4\pi}$
distributions at masses between 1.9 and 2.5~$\gev/c^2$ and 1.4 and
2.2~$\gev/c^2$, respectively. Possible structure within these
broad peaks cannot be resolved with the current statistics. No obvious
structure is observed in other final states. The branching fractions
for $m_{hs} < 2.9~\gev/c^2$ in this paper sum up to
0.26\%~\cite{note1} of the total $\psip$ decay width, which is about a
quarter of the total expected radiative $\psip$ decays.  This
indicates that a larger data sample is needed to search for more decay
modes and to resolve the substructure of $\psip$ radiative decays.

Table~\ref{Br-pi0bg} lists the results of $\psip$ decays into $\piz$ +
hadrons together with the world averaged values~\cite{PDG}, and values
of $Q_h~[=\BR(\psipto h)/\BR(\jpsito h)]$.  For $\psipto\piz 2(\pp)$
decay, intermediate resonances including $\sigma~[f_0(600)]$,
$f_2(1270)$, $\omega$, and $b_1(1235)$ are observed, and the measurement
of $\BR[\psipto\omega f_2(1270)]$ agrees with the previous measurement
using the same data sample~\cite{bes2VP}.  The $\rho^\pm$ resonance is
observed in $\psipto\piz\kskp$ decay mode.
\begin{table*}
\begin{center}
\caption{\label{Tot-nev} Results for $\psipto\gamma +hadrons$. For
each final state, the following quantities are given: the number of
events in $\psip$ data, $N^{Tot}$; the number of background events
from $\psip$ decays and continuum, $N^{Bg}$; the number of signal
events,  $N^{Sig}$; and the weighted averaged efficiency, $\eff$;
the branching fraction with statistical and systematic errors or the
upper limit on the branching fraction at the 90\% C.L. For all the
radiative channels, except the $\gamma\ppb\piz$ and
$\gamma\piz2(\pip\pim)\kap\kam$ modes we require
$m_{hs}<2.9\gev/c^2$. The branching fraction for
$\gamma\piz2(\pip\pim)\kap\kam$ is measured with the requirement
$m_{\piz2(\pip\pim)\kap\kam}<3.38$ GeV/$c^2$. Possible interference
effects for the modes with intermediate states are ignored. }

\begin{tabular}{cccccc} \hline \hline
Mode & $N^{Tot}$ & $N^{Bg}$ & $N^{Sig}$ & $\eff$(\%) &
$\BR(\times 10^{-5})$\\\hline
$\gamma\ppb$ & $329$ & $187$ & $142\pm18$ & 35.3 & 2.9$\pm$0.4$\pm$0.4 \\
$\gamma\ppb\piz$ & $345$ & $219$ & $126 \pm 38$ & 8.94 & $10.1\pm3.1\pm2.1$\\
$\gamma 2(\pi^+\pi^-)$ & $1697$ & $1114$ & $583\pm41$ & 10.4  & 39.6$\pm$2.8$\pm$5.0\\
$\gamma\ks\kap\pim +c.c.$  & $-$ & $-$ & $115\pm16$ & 4.83 & 25.6$\pm$3.6$\pm$3.6 \\
$\gamma K^+ K^-\pi^+\pi^-$ &$361$ &$229$  &$132\pm19$ & 4.94   & 19.1$\pm$2.7$\pm$4.3 \\
$\gamma K^{*0} K^+\pi^-+c.c.$&$-$ &$-$ & $237\pm39$ & 6.86 & 37.0$\pm$6.1$\pm$7.2\\
$\gamma \kstarz\kstarzb$&$58$&$17$&$41\pm8$&2.75& 24.0$\pm$4.5$\pm$5.0\\
$\gamma \pip\pim\ppb$& $55$ & $38$ & $17\pm7$ &4.47 & 2.8$\pm$1.2$\pm$0.7 \\
$\gamma2(\kap\kam)$ & $15$ & $8$  & $<14$ & 2.93& $<4.0$\\
$\gamma3(\pp)$& $118$ & $95$ & $<45$& 1.97 & $<17$\\
$\gamma2(\pi^+\pi^-)K^+K^-$&$17$ & $13$ & $<15.5$ & 0.69 & $<22$ \\
$\gamma\piz2(\pip\pim)\kap\kam$ & $27$ & $21$ & $<24.9$ & $0.058$ & $<310$\\
\hline \hline
\end {tabular}
\end{center}
\end{table*}
\begin{table*}
\begin{center}
\caption{\label{Br-pi0bg} Results of $\psipto\piz +hadrons$. Here
$N^{Sig}$ is the number of signal events, $\eff$ is the detection
efficiency, $\BR$ is the measured branching fraction,
$\BR^{\textrm{PDG}}$ is the world averaged value~\cite{PDG}, and
$Q_h =\BR(\psipto h)/\BR(\jpsito h)$. The branching fraction for
$\piz2(\pip\pim)\kap\kam$ is measured with the requirement
$m_{2(\pip\pim)\kap\kam}<3.38$ GeV/$c^2$.}
\begin{tabular}{cccccccc} \hline \hline
Mode: $h$ & $N^{Sig}$ & $\eff$(\%) & $\BR(\times 10^{-4})$ &
$\BR^{\textrm{PDG}} (\times 10^{-4})$& $Q_h$(\%)\\ \hline
$\piz2(\pip\pim)$ & $2173\pm53$ & $6.32$ & $24.9\pm0.7\pm3.6$&$23.7\pm
2.6$&$10.5\pm2.0$\\
$\omega\pip\pim$ & $386\pm23$ & $3.74$ & $8.4\pm0.5\pm1.2$&$6.6\pm1.7$&$11.7\pm2.4$\\
$\omega f_2(1270)$ & $57\pm13$& $3.65$ & $2.3\pm0.5\pm0.4$ &$2.0\pm0.6$&$5.4\pm0.6$\\
$b_1^\pm\pi^\mp$& $202\pm21$ & $3.24$ & $5.1\pm0.6\pm0.8$
&$3.6\pm0.6$&$17.0\pm4.2$ \\
$\ppb\piz\piz$ & $203 \pm 27$ & 8.30 & $1.75\pm0.21\pm0.28$ &---&---\\
$\piz\ks K^\pm\pi^\mp$&$361\pm25$&4.40&$8.9\pm0.6\pm1.1$&---&---\\
$K^\pm\rho^\mp\ks$&$100\pm20$&3.80&$2.9\pm0.6\pm0.4$&---&---\\
$\piz2(\pip\pim)\kap\kam$ & $48\pm11$ & $0.41$ & $8.4\pm1.9\pm1.7$ &---&---\\
\hline \hline
\end {tabular}
\end{center}
\end{table*}

In summary, we report measurements of the branching fractions of
$\psip$ decays into $\gamma\ppb$,
  $\gamma 2(\pip\pim)$, $\gamma \kskp$, $\gamma \kap\kam\pip\pim$,
  $\gamma\kstarz\kam\pip+c.c.$, $\gamma \kstarz \kstarzb$,
  $\gamma\pip\pim\ppb$, $\gamma2(\kap\kam)$, $\gamma3(\pip\pim)$,
  $\gamma2(\pip\pim)\kap\kam$ and the differential branching
fractions for $\psip$ decays into $\gamma\ppb$, $\gamma
2(\pip\pim)$, $\gamma\kap\kam\pip\pim$, and $\gamma \kskp$
  with hadron invariant mass less than 2.9$\gev/c^2$.
We
  also report branching fractions of $\psip$ decays into $\gamma\ppb\piz$,
  $\ppb\piz\piz$, $\piz\kskp$, $K^\pm\rho^\mp\ks$,
  $\piz2(\pip\pim)\kap\kam$ and $\gamma\piz2(\pip\pim)\kap\kam$.
 The
measurements of $\psip$ decays into $\piz2(\pip\pim)$,
$\omega\pip\pim$, $\omega f_2(1270)$, and $b_1^\pm\pi^\mp$ are
consistent with previous measurements~\cite{PDG} and the recent
measurements by the CLEO collaboration~\cite{chic23hs}.

\acknowledgments The BES collaboration thanks the staff of BEPC and
computing center for their hard efforts. This work is supported in
part by the National Natural Science Foundation of China under
contracts Nos. 10491300, 10225524, 10225525, 10425523, 10625524,
10521003, 10775142, the Chinese Academy of Sciences under contract
No. KJ 95T-03, the 100 Talents Program of CAS under Contract Nos.
U-11, U-24, U-25, and the Knowledge Innovation Project of CAS under
Contract Nos. U-602, U-34 (IHEP), the National Natural Science
Foundation of China under Contract No. 10225522 (Tsinghua
University), and the Department of Energy under Contract No.
DE-FG02-04ER41291 (U. Hawaii).


\begin{thebibliography}{**}
\bibitem{Jdecay}  L. K$\ddot{\hbox{o}}$pke and N. Wermes,
             Phys. Rep. {\bf 174}, 67 (1989).
\bibitem{QWG} N. Brambilla \etal,  hep-ph/0412158.
\bibitem{PDG}W.-M. Yao \etal, Journal of Physics {\bf G 33}, 1 (2006).
\bibitem{PRD-wangp} P. Wang, C. Z. Yuan, and X. H. Mo, Phys. Rev. {\bf D 70}, 114014 (2004).
\bibitem{prlrad} BES Collaboration, M. Ablikim \etal, Phys. Rev.
Lett. {\bf 99}, 011802 (2007).
\bibitem{bes} BES Collaboration, J. Z. Bai \etal, Nucl. Instr. Meth.
              {\bf A 344}, 319 (1994).
\bibitem{bes2}BES Collaboration, J. Z. Bai \etal, Nucl. Instr. Meth. {\bf A  458}, 627 (2001).
\bibitem{pspscan} X. H. Mo {\em et al.}, HEP\&NP {\bf 28}, 455 (2004) [arXiv:hep-ex/0407055].
\bibitem{lum} S. P. Chi {\em et al.}, HEP \& NP \textbf{28}, 1135 (2004).
\bibitem{simbes}BES Collaboration, M. Ablikim \etal,
Nucl. Instrum. Meth. {\bf A 552}, 344 (2005).
\bibitem{chenjc} J. C. Chen {\em et al.}, Phys. Rev. {\bf D 62}, 034003 (2000).
\bibitem{fitchi2} R. G. Ping {\em et al.}, HEP \& NP {\bf 31}, 229 (2007)
  [arXiv:physics/0608213].
\bibitem{ppbpi0-bes2}BES Collaboration, M. Ablikim \etal, Phys. Rev. {\bf D 71}, 072006 (2005).
\bibitem{jpsi-gppb}BES Collaboration, M. Ablikim \etal,  Phys. Rev. Lett. {\bf 91}, 022001 (2003).
\bibitem{massres} The $\ppb$ mass resolution in
the fitted region is less than $3\mev/c^2$ and neglected
in the fit.
\bibitem{r-aston} D. Aston \etal, Nucl. Phys. {\bf B 296}, 493 (1988).
\bibitem{chic23hs} CLEO Collaboration, R. A. Briere \etal, Phys. Rev. Lett. {\bf 95}, 062001 (2005).
\bibitem{pole} J. Conrad, O. Botner, A. Hallgren, and C. Perez de los Heros, Phys. Rev. {\bf D 67}, 012002 (2003).
\bibitem{g4pi} DM2 Collaboration, D. Bisello \etal, Phys. Rev. {\bf D 39}, 701 (1989);
MARK-III Collaboration, R. M. Baltrusaitis \etal,  Phys. Rev. {\bf D
33}, 1222 (1986).
\bibitem{note1} This value includes the decays of
$\psipto\gamma\pip\pim\piz\piz$, $\gamma\ks\kap\pim + c.c.$; the
intermediate resonance channels, e.g. $\psipto\gamma\kstarz\kstarzb$ are excluded.
\bibitem{bes2VP}  BES Collaboration, M. Ablikim \etal, Phys. Rev. {\bf D 69}, 072001 (2004).
\end{thebibliography}
\end{document}